\documentclass[aps,prfluids,preprintnumbers,superscriptaddress]{revtex4-2}
\usepackage[T1]{fontenc}
\usepackage[utf8]{inputenc}
\usepackage{amsmath,amssymb,amsfonts}
\usepackage{graphicx,subcaption}
\usepackage{booktabs,tabularx,supertabular}
\usepackage{microtype}
\usepackage{placeins}
\usepackage[table]{xcolor}
\usepackage[colorlinks=true,linkcolor=blue,citecolor=blue,urlcolor=blue]{hyperref}
\graphicspath{{./}{fig/}}
\newcommand{\dd}{\mathrm{d}}

\begin{document}
\title{Yielding-dependent Taylor dispersion in electrokinetic--magnetohydrodynamic Casson microflows}
\author{Nanda Poddar}
\email{nandap@srmist.edu.in}
\affiliation{Department of Mathematics, SRM Institute of Science and Technology, Kattankulathur, Chennai, Tamil Nadu 603203, India}
\author{Debabrata Das}
\affiliation{Department of Mathematics, Indian Institute of Technology Hyderabad, Hyderabad 502285, India}
\author{Suman Chakraborty}
\email{suman@mech.iitkgp.ac.in}
\affiliation{Department of Mechanical Engineering, Indian Institute of Technology Kharagpur, Kharagpur 721302, India}
\begin{abstract}
Longitudinal dispersion in a yielding fluid depends on the velocity distribution and on how the flow is maintained. We study a neutral tracer in pressure-assisted electrokinetic and magnetohydrodynamic flow of a Casson fluid through a slit microchannel. The nonlinear stress and velocity equations determine the yielded layers, the unyielded core and the axial flow. Mei's multiple-scale homogenization of the advection--diffusion equation then separates transverse relaxation, mean drift and longitudinal spreading, yielding the Taylor--Aris auxiliary problem and effective diffusivity. An exact pressure-driven solution gives cubic decay of the mean velocity and eighth-power decay of the dispersion enhancement near flow arrest. Under fixed pressure and electrical forcing, increasing the yield parameter from $Y=0$ to $0.7$ moves the optimum Hartmann number from $Ha_p=0.696$ to $2.634$, while the corresponding normalized diffusivity decreases from $D_T^{\mathrm C}=1.454$ to $1.015$. At fixed mean throughput and $Ha_p=1$, the trend reverses: $D_T^{\mathrm C}$ increases from $1.000433$ to $1.002311$. The homogenized energy identity traces this difference to the velocity departure from the sectional mean, including that within the unyielded core. Analytical limiting solutions, direct concentration calculations, numerical convergence studies and a coupled Casson shear test support the reduced-flow predictions. An electrolyte-based dimensional assessment identifies the magnetic optima as theoretical forcing states for representative aqueous microchannels. These results are relevant to predicting longitudinal tracer spreading in microchannels carrying fluids with a yield stress, whether pressure or throughput is controlled.
\end{abstract}
\maketitle

	\section{Introduction}
The longitudinal spreading of a solute in channel flow arises from the interplay of axial advection and transverse molecular diffusion. Taylor~\cite{ta59} and Aris~\cite{ar56} established the effective axial diffusivity that emerges after transverse equilibration. Mei and coworkers~\cite{mei96,mei10} developed a multiple-scale homogenization procedure in which solvability conditions successively determine the mean drift and the effective dispersion. Transient and moment-based descriptions further characterize the approach to this limit~\cite{Chatwin1970,GillSankarasubramanian1970,Barton1983,VedelHovadBruus2014,wu14}. Yield-stress flow raises a particular question for this framework: solute continues to diffuse through an unyielded region even though the local shear vanishes. The full cross-sectional velocity, including the plug, must therefore enter the dispersion calculation.

Electroosmosis offers an alternative to pressure-driven transport in narrow channels. An imposed tangential electric field acts on the electrical double layer (EDL) adjacent to charged walls, generating a body force whose spatial distribution depends on the ionic environment and surface potential \cite{gr99,li05,paul12,Reshadi19,Sadeghi20,ro23,hu24}. Electrokinetic transport is used in microfluidic sample handling, where axial band broadening can affect the distribution of dissolved species~\cite{st04,ib13,sa15}. For sufficiently thin EDLs, electroosmosis can produce a nearly uniform bulk velocity and thereby reduce shear-induced spreading. When an electric current and magnetic field are applied simultaneously, the Lorentz force offers a further means of changing the axial momentum balance. Electromagnetohydrodynamic (EMHD) flows and magnetically modified dispersion in microchannels have consequently attracted considerable attention \cite{ch06,ts01,mu09,da12,ya16,si15,bu14,es14,Vargas2017}.

Non-Newtonian rheology introduces another important element into this coupling. Solute dispersion in complex and viscoplastic fluids has been examined in pressure-driven, oscillatory and electrokinetic settings \cite{nag04,nag09,ran16,ra17,ro18,ro21,das21}. More general layered-fluid and reactive-transport settings provide complementary context \cite{de17,de17a}. In particular, Dash \emph{et al.}~\cite{DashJayaramanMehta2000} investigated the influence of Casson yield stress on longitudinal solute dispersion in channels and tubes, including comparisons under equivalent flow conditions. In a yield-stress fluid, the stress may be insufficient to produce local shearing, giving rise to unyielded (locally shear-free) regions whose boundaries must be found as part of the flow solution \cite{Papanastasiou1987,BalmforthFrigaardOvarlez2014}. Ng~\cite{Ng2013Slit} analyzed pressure-assisted electroosmotic Casson flow in a slit, including yielded and unyielded layers, while Ng and Qi~\cite{Ng2013} considered electroosmotic motion of viscoplastic liquids. The location of the yield surfaces and the velocity carried by the plug are consequently as important to solute spreading as the shear in the adjoining layers.

Recent work has examined electrically and magnetically controlled solute spreading in Newtonian and complex fluids. Das \emph{et al.}~\cite{Das2024} studied magnetic modulation of pulsatile electroosmotic transport, Sen \emph{et al.}~\cite{Sen2025} investigated electromagnetic dispersion in a couple-stress liquid, and Hossain \emph{et al.}~\cite{Hossain2025} reported a nonmonotonic magnetic response in electroosmotic flow. Casson transport studies have treated electroosmotic dispersion~\cite{Debnath2025}, multiscale transport with wall absorption~\cite{AyyanarRaghavanBarik2026}, and pulsatile flows with interphase exchange~\cite{RanaRoyDebnathPatne2026}. Sen and Kairi~\cite{SenKairi2026} considered an electroosmotic--EMHD Casson channel with buoyancy, slip-dependent surface potential and wall reactions. Further electromagnetic and couple-stress configurations appear in Refs.~\cite{GuptaKundu2026,DasDharKairiMondal2025}. Boyko and Stone~\cite{BoykoStone2026} examined Taylor--Aris dispersion in shear-rate-dependent flows, including the role of tracer diffusivity. Here the tracer diffusivity is constant; the rheological dependence enters through the computed Casson velocity and the positions of the yield surfaces.

The interaction of yielding with electromagnetic forcing raises two related questions: how do moving Casson yield surfaces affect the magnetic dispersion maximum, and does the response persist when mean throughput replaces imposed forcing as the operating constraint? Under fixed forcing, both the velocity profile and the flow rate change with yield stress; at fixed throughput, the pressure gradient adjusts to sustain the prescribed mean velocity. The homogenized dispersion problem allows these effects to be compared on the same basis. The distinction matters when interpreting sample-band spreading in pressure-controlled and flow-rate-controlled microchannels. We address these questions in the fully developed slit-flow limit.

We determine the Casson stress, velocity and yield surfaces simultaneously, then use Mei's multiple-scale expansion through the third concentration correction to obtain the axial dispersion and transverse concentration structure. Exact pressure-driven Casson flow and a closed-form constant-viscosity solution provide benchmarks. The dispersion predictions are tested against direct two-dimensional transport, independent moment calculations and a two-component Casson shear model. We also examine the dimensional electrolyte constraints on the magnetic optima.

Section~II develops the electrokinetic and Casson flow equations, followed by the multiple-scale derivation of the axial transport law and the numerical procedures. Section~III establishes analytical benchmarks, examines yielding-dependent magnetic dispersion under fixed forcing and fixed throughput, and tests the predictions against finite-time transport and transverse-shear calculations. The electromagnetic parameter constraints are assessed in the same section. Section~IV summarizes the physical conclusions. Appendices~\ref{app:multiscale}, \ref{app:coefficients}, \ref{app:neumann_operator} and \ref{app:near_arrest} contain, respectively, the multiple-scale solvability calculations, the linear-viscosity dispersion solution, the inverse Neumann operator, and the near-arrest expansion.

\section*{Nomenclature}
\renewcommand{\arraystretch}{1.02}
\footnotesize
\setlength{\tabcolsep}{5pt}
\tablefirsthead{}
\tablehead{%
\hline
\multicolumn{4}{c}{\emph{Nomenclature (continued)}} \\
\hline
\textbf{Symbol} & \textbf{Description} & \textbf{Symbol} & \textbf{Description} \\
\hline
}
\begin{supertabular}{ll@{\hspace{0.6cm}}ll}
\hline
\textbf{Symbol} & \textbf{Description} &
\textbf{Symbol} & \textbf{Description} \\
\hline
$\langle C\rangle$ & Cross-sectional mean of full concentration                          & $\mu$         & Dynamic viscosity ($\mathrm{Pa\,s}$) \\
$A_1,A_2,A_3$& Linear-model auxiliary functions                                & $Y$         & Dimensionless yield-stress parameter \\
$B_0$        & Applied magnetic field (T)                         & $\mu_p$       & Casson plastic viscosity \\
$\langle\cdot\rangle$ & Cross-sectional average operator                              & $\rho$        & Fluid density ($\mathrm{kg\,m^{-3}}$) \\
$C$          & Full dimensionless solute concentration                 & $\rho_e$      & Net charge density ($\mathrm{C\,m^{-3}}$) \\
$C^{*}$      & Dimensional solute concentration                   & $\sigma$      & Electrical conductivity ($\mathrm{S\,m^{-1}}$) \\
$C_0$        & Leading-order homogenized concentration                            & $\tau$        & Moving-frame dimensionless time \\
$C_1,C_2,C_3$& Perturbation concentration terms                   & $\boldsymbol{\tau}$ & Stress tensor \\
$D$          & Molecular diffusion coefficient                    & $\psi$        & Dimensionless electric potential \\
$D_T$        & $D_{\mathrm{eff}}/D$, total normalized diffusivity                      & $\psi^{*}$    & Electric potential (V) \\
$e$          & Elementary charge                                 & $\zeta$       & Zeta potential (V) \\
$D_T^*$      & Shear-induced dispersion factor (linear model)                              & $\beta$       & Linear-viscosity parameter \\
$E_x$        & Axial electric field ($\mathrm{V\,m^{-1}}$)        & $\beta_1$     & Pressure-to-electroosmotic velocity ratio \\
$E_z$        & Spanwise electric-field magnitude ($\mathrm{V\,m^{-1}}$)   & $\delta(\cdot)$ & Dirac delta function \\
$F$          & Electromagnetic body force per unit volume         & $\epsilon$    & Fluid permittivity ($\mathrm{F\,m^{-1}}$) \\
$h$          & Half channel height (m)                            & $\varepsilon$ & Perturbation parameter ($h/L$) \\
$Ha$         & Hartmann number                                    & $D_T^{\mathrm C}$ & Casson normalized diffusivity \\
$J$          & Current density ($\mathrm{A\,m^{-2}}$)             & $\kappa$      & Electrokinetic parameter \\
$k_B$        & Boltzmann constant                                 & $\lambda_D$   & Debye length (m) \\
$L$          & Channel length (m)                                 & $\xi$         & Moving coordinate \\
$n_{+},n_{-}$& Number densities of cations and anions             & EDL           & Electrical double layer \\
$n_0$        & Bulk number density of each ion species                           & EOF           & Electroosmotic Flow \\
$P$          & Dimensional pressure-gradient forcing (Pa/m)                        & EMHD          & Electro-Magnetohydrodynamics \\
$p^*$        & Dimensional pressure (Pa)                                      & MHD           & Magnetohydrodynamics \\
$Pe$         & Electroosmotic P\'eclet number                                      & $Q$           & Volumetric flow rate \\
$\tau_y$     & Casson yield stress                               & $x^*,y^*,z^*$      & Dimensional Cartesian coordinates  \\
$S$          & Linear-model spanwise-field parameter  & $t,t^*$     & Dimensionless/dimensional times \\
$t_0,t_1,t_2$& Multiple time scales                               & $T_{av}$      & Average absolute temperature \\
$u$          & Dimensionless axial velocity                       & $u^{*}$       & Dimensional axial velocity \\
$u_{HS}$     & Helmholtz--Smoluchowski velocity                   & $u_p$         & Pressure-driven characteristic velocity \\
$U$          & Velocity vector                                    & $W$           & Channel width (m) \\
$x,y$      & Dimensionless axial/wall-normal coordinates                              & $z$           & Ionic valence \\
$\alpha$ & Yield-to-wall-stress ratio & $\delta$ & Distance to flow arrest ($1-\alpha$) \\
$\Pi$ & Dimensionless pressure gradient & $\Omega$ & Dimensionless electroosmotic forcing \\
$Ha_p$ & Plastic-viscosity Hartmann number & $S_p$ & Spanwise-field parameter based on $\mu_p$ \\
$f_{\mathrm u}$ & Fraction of slit remaining unyielded & $y_f$ & Upper-half-channel yield-front position \\
$Pe_0$ & Reference-velocity P\'eclet number & $\widehat A_1$ & Casson first-order auxiliary function \\
$U_0$ & Reference velocity in yielding model & $U_P$ & Pressure-driven velocity scale \\
$G$ & Pressure-gradient magnitude ($\mathrm{Pa\,m^{-1}}$) & $D_{\mathrm{eff}}$ & Dimensional effective axial diffusivity \\
$D_+,D_-$ & Ionic diffusivities & $J_z$ & Spanwise electric-current density \\
$\widehat{u},\widehat{\tau}$ & Dimensionless velocity and shear stress & $m$ & Effective magnetic screening parameter \\
$Ha_{\mathrm{cr}}$ & Hartmann number at dispersion extremum & $I$ & Relative dispersion increment (\%) \\
$V$ & Transverse concentration-range measure & $d_1,\ldots,d_4$ & Linear-model analytical coefficients \\
$\overline{u}$ & Dimensional mean axial velocity & $\overline D_i$ & Mean ionic diffusivity \\
$a(y)$ & Pressure-driven auxiliary function & $\epsilon_r$ & Relative permittivity \\
$\epsilon_0$ & Vacuum permittivity & $\mu_{\mathrm{eff}}$ & Effective viscosity in the linear law \\
$B_{\mathrm{req}}$ & Field needed to reach a reference $Ha_p$ & $\sigma_0$ & Initial Gaussian pulse standard deviation \\
$U_{\mathrm{ref}},Pe_{\mathrm{ref}}$ & Pressure-reference velocity and P\'eclet number & $\Pi_{\mathrm{ref}}$ & Pressure parameter in electrically unforced reference \\
$q=\xi/Pe$ & Rescaled axial coordinate (electrically forced cases) & $\Delta A_1$ & Range of first auxiliary function \\
\hline
\end{supertabular}
\normalsize

\section{Mathematical formulation}

	\subsection{Channel geometry and imposed fields}
\begin{figure}
	\centering
	\begin{subfigure}{0.37\linewidth}
		\centering
		\includegraphics[width=\linewidth]{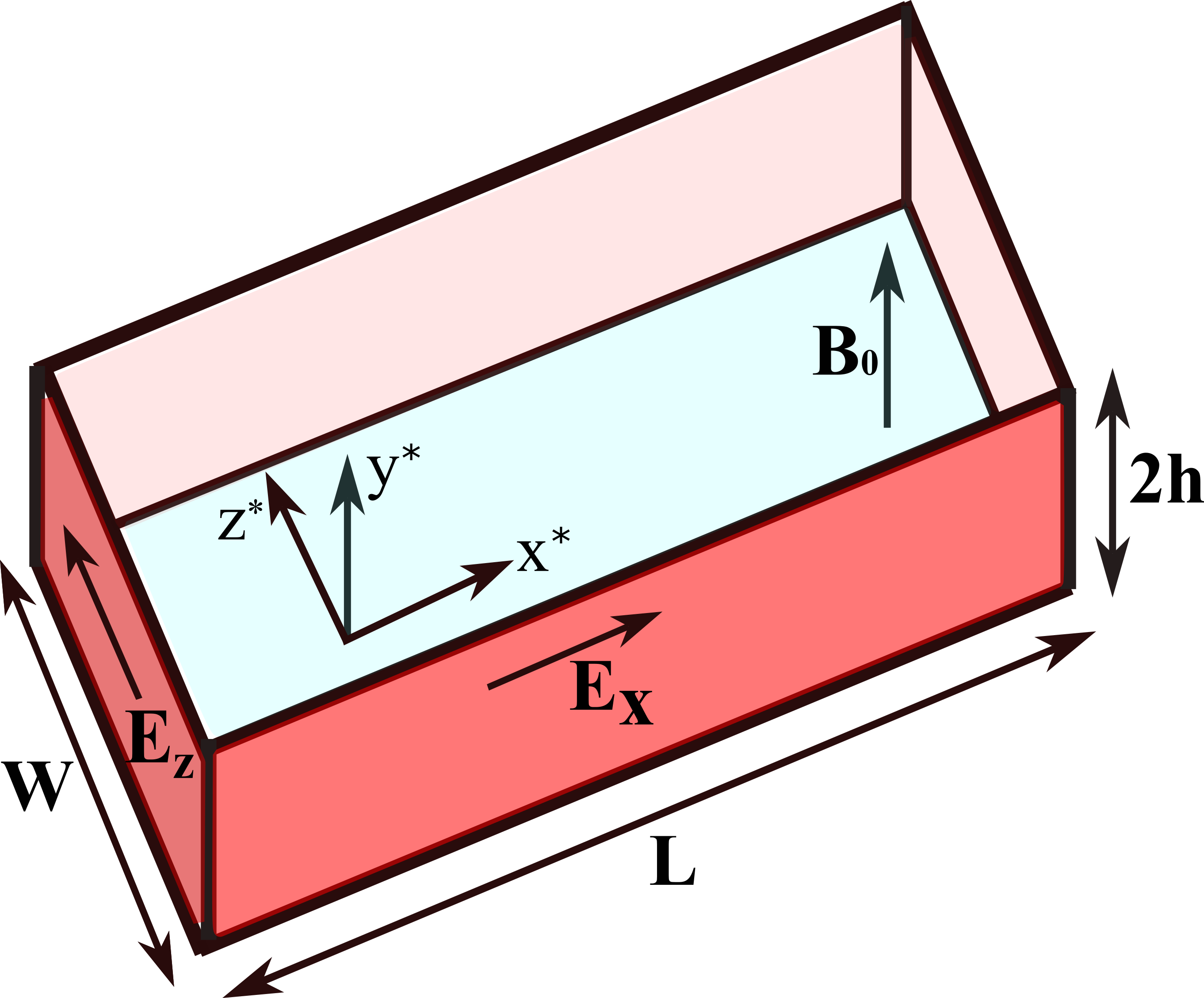}
		\caption{}
		\label{fig:1a}
	\end{subfigure}
	\begin{subfigure}{0.6\linewidth}
		\centering
		\includegraphics[width=\linewidth]{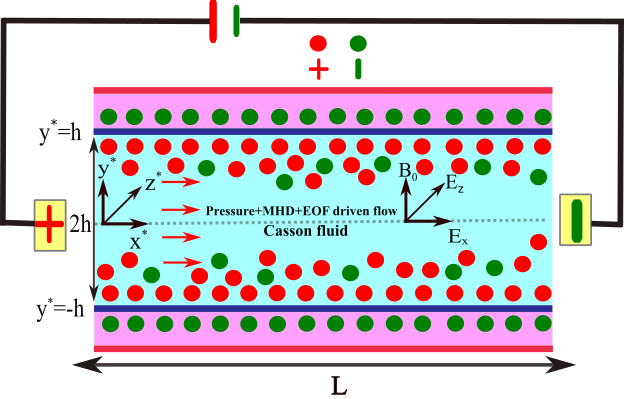}
		\caption{}
		\label{fig:1b}
	\end{subfigure}
	\caption{Slit-microchannel geometry and imposed electrokinetic--magnetohydrodynamic forcing.}
	\label{fig:1}
\end{figure}

Consider a steady, incompressible flow of a Casson liquid through a slit microchannel of height $2h$, width $W$ and length $L$, with $2h\ll W$ and $h\ll L$. The coordinates $(x^*,y^*,z^*)$ are directed along the channel, normal to its walls and across its width. As shown in Fig.~\ref{fig:1}, an axial pressure gradient and electric field $E_x$ drive the flow; a magnetic field $B_0$ acts in the wall-normal direction, and an externally maintained electric-field component $-E_z$ acts spanwise. The corresponding axial Lorentz force density is $\sigma B_0 E_z-\sigma B_0^2u^*$. Both walls have zeta potential $\zeta$. The flow is taken to be fully developed and unidirectional, $\mathbf U=u^*(y^*)\mathbf e_x$, where $u^*$ is the dimensional axial velocity and $\mathbf e_x$ is the streamwise unit vector. An external circuit and side electrodes are assumed to maintain the imposed spanwise electric-field component. The axial model is uniform in the spanwise direction and omits flow near the side electrodes. The imposed field also drives charge within the EDL, which can generate spanwise shear and alter the Casson yield surfaces. We therefore treat the axial shear law as an approximation for regimes in which this coupling has little effect on the axial flow. Section~\ref{sec:transverse_sensitivity} tests its accuracy with a two-component Casson calculation. The electrical current and pressure conditions of a finite-width device require a separate closure; in particular, an imposed spanwise current is incompatible with insulating side walls unless an appropriate return path is provided. A dilute neutral tracer is released at $t^*=0$ across the channel section at $x^*=0$. The large-width approximation and spanwise-uniform release reduce the transport problem to the $(x^*,y^*)$ plane.

\subsection{Electric double-layer potential}

Under the equilibrium-EDL approximation for a dilute electrolyte \cite{li05,gr99,Ng2013}, the electric potential $\psi^*$ satisfies Poisson's equation
\begin{equation}\label{eq:1}
	\nabla^2 \psi^*=-\frac{\rho_e}{\epsilon},
\end{equation}
Here $\rho_e$ denotes the local ionic charge density and $\epsilon$ is the permittivity of the electrolyte. For a symmetric $z:z$ electrolyte, the charge density is

\begin{equation}\label{eq:2}
	\rho_e=ez(n_+-n_-),
\end{equation}
Here $z$ is the ionic valence and $n_+$ and $n_-$ are the local number densities of cations and anions. Assuming dilute, ideal ions in thermal equilibrium, their number densities obey the Boltzmann distributions

\begin{equation}\label{eq:3}
	n_\pm= n_0\, \exp\left(\mp\frac{ze\psi^*}{k_B \,T_{av}}\right),
\end{equation}
In Eq.~\eqref{eq:3}, $n_0$ is the bulk number density of either ion species, $e$ the elementary charge, $k_B$ the Boltzmann constant and $T_{av}$ the absolute temperature. For sufficiently small wall potentials, $|ze\zeta/(k_BT_{av})|\ll1$, the Debye--H\"uckel linearization and the scalings $\psi=\psi^*/\zeta$ and $y=y^*/h$ give
\begin{equation}\label{eq:4}
	\frac{\dd^2 \psi}{\dd y^2}=\kappa^2\, \psi,
\end{equation}
with boundary conditions,
\begin{equation}\label{eq:5}
	\psi=1\qquad\mbox{at}\qquad y=\pm 1,
\end{equation}
The Debye length is $\lambda_D=(ez)^{-1}[\epsilon k_BT_{av}/(2n_0)]^{1/2}$, and $\kappa=h/\lambda_D$ measures the relative thickness of the EDL.

The solution of Eqs.~\eqref{eq:4} and \eqref{eq:5} is
\begin{equation}\label{eq:6}
	\psi=\frac{\cosh(\kappa y)}{\cosh(\kappa)}.
\end{equation}

The corresponding ionic charge density follows from Eq.~\eqref{eq:2}:
\begin{equation}\label{eq:7}
	\rho_e=-\frac{\kappa^2\,\epsilon\,\zeta}{h^2}\frac{\cosh(\kappa y)}{\cosh(\kappa)}.
\end{equation}
\subsection{Casson rheology and yield surfaces}

For a Casson material, let $\mu_p$ denote the plastic-viscosity parameter and $\tau_y\geq0$ the yield stress. The dimensional stress component $\tau_{xy}$ is the shear traction acting in the streamwise direction across a wall-normal surface. The exact constitutive law in simple shear has the inverse form \cite{Ng2013,BalmforthFrigaardOvarlez2014}
\begin{equation}\label{eq:8}
\frac{\dd u^*}{\dd y^*}=\begin{cases}
0, & |\tau_{xy}|\leq\tau_y,\\[3pt]
\displaystyle\frac{\operatorname{sgn}(\tau_{xy})}{\mu_p}
\left(\sqrt{|\tau_{xy}|}-\sqrt{\tau_y}\right)^2,
& |\tau_{xy}|>\tau_y.
\end{cases}
\end{equation}
The strain rate is zero wherever $|\tau_{xy}|\leq\tau_y$. Such an \emph{unyielded region} translates without local shear as a rigid plug, whereas a \emph{yielded region}, characterized by $|\tau_{xy}|>\tau_y$, supports deformation according to the nonlinear constitutive law. Their interface is a yield surface at $|\tau_{xy}|=\tau_y$ and must be determined from the stress and wall conditions. The presence of a plug does not suppress molecular diffusion of the dissolved tracer.

For comparison, consider a linear constitutive relation with an effective viscosity independent of the local shear rate,
\begin{equation}\label{eq:9}
\tau_{xy}=\mu\left(1+\frac{1}{\beta}\right)\frac{\dd u^*}{\dd y^*},
\end{equation}
where $\beta>0$ controls the constant effective viscosity $\mu_{\mathrm{eff}}=\mu(1+1/\beta)$. The linear law has no yield surface and serves as an analytical reference for separating the electromagnetic response from the effects of Casson yielding. In the limit $\beta\to\infty$, the viscosity approaches $\mu$. The Casson law~\eqref{eq:8} independently tends to Newtonian behavior as $\tau_y\to0$, with $\mu_p$ fixed.

\subsection{Fully developed momentum balance}

The momentum balance is

\begin{equation}\label{eq:10a}
	\rho\frac{D\mathbf U}{Dt}=-\nabla p^*+\nabla\cdot\boldsymbol{\tau}+\mathbf F,	
\end{equation}

Here $p^*$ denotes dimensional pressure, $\mathbf U$ the fluid velocity, $\boldsymbol\tau$ the deviatoric stress tensor, $\mathbf F$ the electromagnetic body-force density, and $\rho$ the fluid mass density. The material derivative is denoted by $D/Dt$. In the low-Reynolds-number and low-magnetic-Reynolds-number regime, fluid inertia and magnetic induction are neglected. With $\mathbf B=B_0\mathbf e_y$, $\mathbf E=E_x\mathbf e_x-E_z\mathbf e_z$ and $\mathbf U=u^*(y^*)\mathbf e_x$, where $\mathbf e_y$ and $\mathbf e_z$ are the wall-normal and spanwise unit vectors and $\sigma$ is electrical conductivity, the quasi-static Ohm law gives
\begin{equation}\label{eq:ohms}
\mathbf J=\sigma(\mathbf E+\mathbf U\times\mathbf B),\qquad
J_z=\sigma(u^*B_0-E_z),
\end{equation}
and thus $(\mathbf J\times\mathbf B)_x=\sigma B_0E_z-\sigma B_0^2u^*$. Under the fully developed approximation, the \emph{general} axial momentum equation is
\begin{equation}\label{eq:momentum_general}
0=-\frac{\dd p^*}{\dd x^*}+\frac{\dd\tau_{xy}}{\dd y^*}
-\sigma B_0^2u^*+\rho_eE_x+\sigma B_0E_z.
\end{equation}
The quasi-static axial projection presumes that the external circuit maintains $E_z$ while accommodating the velocity-induced contribution $u^*B_0$ to $J_z$. It retains the axial electric body force $\rho_e E_x$ and the axial Lorentz force in Eq.~\eqref{eq:momentum_general}. The corresponding spanwise force components include $-\rho_e E_z$ and, within the same Ohmic approximation, $\sigma E_xB_0$. The axial approximation omits these spanwise forces from the velocity calculation. A spanwise velocity $w^*(y^*)$ would introduce a second shear component into the Casson yield criterion, potentially changing the axial velocity and yield surfaces. Accordingly, Eqs.~\eqref{eq:momentum_general}--\eqref{eq:nonlinear_shear} describe the axial projection of the flow; Sec.~\ref{sec:transverse_sensitivity} quantifies the sensitivity to the omitted shear under specified pressure closures.

Substitution of the linear-viscosity relation~\eqref{eq:9} into the general balance~\eqref{eq:momentum_general} gives
\begin{equation}\label{eq:10}
	0=-\frac{\dd p^*}{\dd x^*}+\left(1+\frac{1}{\beta}\right)\mu\frac{\dd^2 u^*}{\dd {y^*}^2}-\sigma B_0^2  u^*+\rho_e E_x+\sigma E_z B_0.
\end{equation}

No slip at the two channel walls gives

\begin{equation}\label{eq:11}
	u^*=0 ~\mbox{at}~ y^*=\pm h.  
\end{equation}

For a constant imposed axial pressure gradient, denote the dimensional pressure forcing $P$ (in $\mathrm{Pa\,m^{-1}}$, of either sign) by
\begin{equation}\label{eq:12}
	-\frac{\dd p^*}{\dd x^*}=P.
\end{equation}

For the effective-viscosity solution of Eq.~\eqref{eq:10}, we introduce
\begin{equation}\label{eq:13}
	y=\frac{y^*}{h}, u=\frac{u^*}{u_{HS}}, Ha=\frac{B_0 h \sqrt{\sigma}}{\sqrt{\mu}},  u_p=\frac{P h^2}{\mu}, \beta_1=\frac{u_p}{u_{HS}}, u_{HS}=-\frac{\epsilon \zeta E_x}{\mu}, S=\frac{E_z h \sqrt{\sigma} }{u_{HS} \sqrt{\mu}} ,
\end{equation}
Here $\beta_1$ is the ratio of the pressure-driven velocity scale $u_p$ to the Helmholtz--Smoluchowski velocity $u_{HS}$, $Ha$ is the Hartmann number based on $\mu$, and $S$ measures the spanwise-electric-field contribution relative to $u_{HS}$. Equation~\eqref{eq:10} becomes

\begin{equation}\label{eq:14}
	\left(1+\frac{1}{\beta}\right)\frac{\dd^2 u}{\dd y^2}- Ha^2u=-\beta_1-S\,Ha-\kappa^2 \frac{\cosh(\kappa y)}{\cosh(\kappa)}.
\end{equation}

The corresponding no-slip conditions are

\begin{equation}\label{eq:15}
	u=0~~\mbox{at}~ y=\pm 1.
\end{equation}

The solution of the linear effective-viscosity momentum equation is

\begin{equation}\label{eq:16}
	u(y)=\frac{\beta_1+\,S\,Ha}{m^2\,\left(1+\frac{1}{\beta}\right)}\left[1-\frac{\cosh(m y)}{\cosh(m)}\right]+\frac{1}{\left(1+\frac{1}{\beta}\right)}\frac{\kappa^2}{\kappa^2-m^2}\left[\frac{\cosh(my)}{\cosh(m)}-\frac{\cosh(\kappa y)}{\cosh(\kappa)}\right],
\end{equation}
where $m^2=Ha^2/\left(1+1/\beta\right)$. Here $m$ is the effective magnetic screening parameter. The apparent singularities at $m=0$ (no magnetic field) and $m=\kappa$ in Eq.~\eqref{eq:16} are removable; the combined expression must be evaluated using the corresponding continuous limits rather than direct substitution.

Equation~\eqref{eq:16} describes the linear-viscosity velocity field. Yielding requires instead the nonlinear constitutive relation~\eqref{eq:8} together with the momentum balance~\eqref{eq:momentum_general}. Taking a characteristic velocity $U_0>0$, the associated shear-stress scale is $\mu_p U_0/h$. Define the dimensionless velocity $\widehat u=u^*/U_0$ and stress $\widehat\tau=h\tau_{xy}/(\mu_p U_0)$, together with
\begin{equation}\label{eq:realparams}
Y=\frac{h\tau_y}{\mu_pU_0},\quad
\Pi=\frac{Ph^2}{\mu_pU_0},\quad
Ha_p=B_0h\sqrt{\frac{\sigma}{\mu_p}},\quad
S_p=\frac{E_zh}{U_0}\sqrt{\frac{\sigma}{\mu_p}},\quad
\Omega=-\frac{\epsilon\zeta E_x}{\mu_pU_0},
\end{equation}
where $Y$ is the yield-stress parameter, $\Pi$ is the dimensionless imposed pressure gradient, $Ha_p$ is the Hartmann number based on plastic viscosity, $S_p$ measures the spanwise electric field, and $\Omega$ is the ratio of the nominal Helmholtz--Smoluchowski velocity to $U_0$. The coupled dimensionless stress--velocity equations are
\begin{align}
\frac{\dd\widehat\tau}{\dd y}
&=-\Pi-\Omega\kappa^2\frac{\cosh(\kappa y)}{\cosh\kappa}
-S_pHa_p+Ha_p^2\widehat u,\label{eq:nonlinear_stress}\\
\frac{\dd\widehat u}{\dd y}
&=\begin{cases}0,&|\widehat\tau|\leq Y,\\[2pt]
\operatorname{sgn}(\widehat\tau)
\left(\sqrt{|\widehat\tau|}-\sqrt Y\right)^2,
&|\widehat\tau|>Y,
\end{cases}\label{eq:nonlinear_shear}
\end{align}
For identical no-slip walls and symmetric driving, the boundary and symmetry conditions are $\widehat u(\pm1)=0$, $\widehat\tau(0)=0$ and $\widehat u(-y)=\widehat u(y)$. Throughout, $\langle f\rangle=\frac12\int_{-1}^{1} f(y)\,\dd y$ denotes the cross-sectional average in the slit. The stress and velocity must be determined simultaneously: the yield surfaces are locations where $|\widehat\tau|=Y$. We write $y_f$ for the yield-front position in the upper half-channel when there is one symmetric central plug. Its dimensionless unyielded fraction is $f_{\mathrm u}=y_f$ in that case; for more general stress distributions, $f_{\mathrm u}=\langle\mathbf{1}_{\{|\widehat\tau|\leq Y\}}\rangle$ denotes the unyielded fraction of the full slit.

For the Casson flow, the P\'eclet number is defined using $U_0$, while the linear-viscosity variables in Eq.~\eqref{eq:22} use the Helmholtz--Smoluchowski velocity $u_{HS}$:
\begin{equation}\label{eq:yielding_pe}
Pe_0=\frac{U_0h}{D},\qquad
\xi=\frac{x^*}{h}-Pe_0\langle\widehat u\rangle\tau,
\qquad \tau=\frac{t^*D}{h^2}.
\end{equation}
For any computed steady axial velocity, the homogenization procedure reduces the dispersive correction to a cross-sectional auxiliary problem. Applied to the Casson profile, the first corrector satisfies
\begin{align}
\frac{\dd^2\widehat A_1}{\dd y^2}
&=\widehat u-\langle\widehat u\rangle,
\qquad \left.\frac{\dd\widehat A_1}{\dd y}\right|_{y=\pm1}=0,
\qquad \langle\widehat A_1\rangle=0,\label{eq:yielding_auxiliary}\\
D_T^{\mathrm{C}}&=\frac{D_{\mathrm{eff}}^{\mathrm{C}}}{D}
=1+Pe_0^2\left\langle
\left(\frac{\dd\widehat A_1}{\dd y}\right)^2\right\rangle.
\label{eq:yielding_dispersion}
\end{align}
In Eq.~\eqref{eq:yielding_dispersion}, the dispersive enhancement is nonnegative and is controlled by the cross-sectional structure of the computed velocity. In particular, an unyielded plug need not have the mean velocity of the channel and can therefore contribute to Taylor dispersion even though its local shear rate is zero. In comparisons of the two constitutive laws, we use the common viscosity scale $\mu=\mu_p$. For $\Omega\neq0$, the dimensionless parameters based on $u_{HS}$ and $U_0$ are related by
\begin{equation}\label{eq:reference_yielding_conversion}
Ha=Ha_p,\qquad
\beta_1=\frac{\Pi}{\Omega},\qquad
S=\frac{S_p}{\Omega},\qquad
Pe=\Omega Pe_0,\qquad
u=\frac{\widehat u}{\Omega}.
\end{equation}
Hence $Pe\,u=Pe_0\widehat u$, and for a fixed \emph{physical} velocity profile the auxiliary functions obey $\widehat A_1=\Omega A_1$. Consequently $1+Pe^2\langle(\dd A_1/\dd y)^2\rangle=1+Pe_0^2\langle(\dd\widehat A_1/\dd y)^2\rangle$ for that same profile. This normalization identity does not equate the predictions of the distinct constitutive laws. The $U_0$ normalization remains well-defined when $\Omega=0$, whereas the electroosmotic scale vanishes. The rheological parameters $Y$ and $\beta$ represent different constitutive properties.

For a pressure-driven comparison in which the imposed axial and spanwise electrical forces are negligible compared with the pressure forcing, the leading-order reference is obtained by setting $E_x=E_z=0$. Normalization by $u_{HS}$ in Eq.~\eqref{eq:13} is undefined. We therefore introduce an independent positive reference velocity $U_{\mathrm{ref}}$, with
\begin{equation}\label{eq:pressure_only_scaling}
  u_{\mathrm{pd}}=\frac{u^*}{U_{\mathrm{ref}}},\qquad
  \Pi_{\mathrm{ref}}=\frac{Ph^2}{\mu U_{\mathrm{ref}}},\qquad
  Pe_{\mathrm{ref}}=\frac{U_{\mathrm{ref}}h}{D}.
\end{equation}
Under the linear-viscosity law, the pressure-driven magnetic-braking reference profile obeys
\begin{equation}\label{eq:pressure_only_velocity}
\left(1+\frac{1}{\beta}\right)\frac{\dd^2 u_{\mathrm{pd}}}{\dd y^2}
-Ha^2u_{\mathrm{pd}}=-\Pi_{\mathrm{ref}},\qquad
u_{\mathrm{pd}}(\pm1)=0,
\end{equation}
with the explicit solution
\begin{equation}\label{eq:pressure_only_exact}
u_{\mathrm{pd}}(y)=\frac{\Pi_{\mathrm{ref}}}{Ha^2}
\left[1-\frac{\cosh(my)}{\cosh m}\right],\qquad
m=\frac{Ha}{\sqrt{1+1/\beta}}.
\end{equation}
The zero-field limit is $u_{\mathrm{pd}}=\Pi_{\mathrm{ref}}(1-y^2)/[2(1+1/\beta)]$. The magnetic damping term is retained because the specified circuit permits the velocity-induced current even when $E_z=0$; insulating side boundaries would require a different current closure. Equations~\eqref{eq:pressure_only_scaling}--\eqref{eq:pressure_only_exact} are used only to interpret the pressure-only comparison panels, with $\Pi_{\mathrm{ref}}=1$ and $Pe_{\mathrm{ref}}=10$. In all formulae for those panels, $Pe_{\mathrm{ref}}$ replaces $Pe$ consistently. The limits $E_x,E_z\to0$ must therefore be taken in dimensional variables or with the independent reference scale; $\beta_1$, $S$ and $Pe$ cannot simultaneously be held fixed under the original $u_{HS}$ normalization.

\subsection{Pressure-driven Casson limit}
An exact limiting solution provides a benchmark for the nonlinear model. With $E_x=E_z=B_0=0$ and $P=G>0$, the shear stress in the upper half-channel is $\tau_{xy}=-Gh\,y$ for $0\le y\le1$. Define the yield-to-wall-stress ratio $\alpha=\tau_y/(Gh)$ and the pressure-driven velocity scale $U_P=Gh^2/\mu_p$. For $0\le\alpha<1$, the central plug occupies $|y|\le\alpha$, and the velocity is
\begin{equation}\label{eq:casson_exact}
\frac{u^*(y)}{U_P}=\begin{cases}
F(\alpha),&0\le|y|\le\alpha,\\[2pt]
F(|y|),&\alpha<|y|\le1,
\end{cases}
\end{equation}
where
\begin{equation}\label{eq:casson_F}
F(y)=\frac{1-y^2}{2}
-\frac{4\sqrt\alpha}{3}(1-y^{3/2})+\alpha(1-y).
\end{equation}
Continuity of $u^*$ and of shear traction holds at the yield surfaces. At $\alpha=0$, Eq.~\eqref{eq:casson_exact} reduces to plane Poiseuille flow; at $\alpha\ge1$, the fluid remains unyielded and no slip implies rest. By symmetry, the exact cross-sectionally averaged velocity is
\begin{align}\label{eq:casson_mean}
\frac{\langle u^*\rangle}{U_P}
&=\int_\alpha^1 r(\sqrt r-\sqrt\alpha)^2\,\dd r\\
&=\frac{1-\alpha^3}{3}-\frac{4\sqrt\alpha}{5}
(1-\alpha^{5/2})+\frac\alpha2(1-\alpha^2).\nonumber
\end{align}
Defining the distance to complete arrest as $\delta=1-\alpha$, for $\delta\to0^+$ with $G$, $h$ and $\mu_p$ fixed,
\begin{equation}\label{eq:nearflux}
\frac{\langle u^*\rangle}{U_P}=\frac{\delta^3}{12}+O(\delta^4).
\end{equation}
The different scaling of longitudinal dispersion in this limit follows from the Taylor--Aris auxiliary boundary-value problem developed below and is derived in Appendix~\ref{app:near_arrest}; it is not assumed to extend to the fully coupled electromagnetic flow.

\subsection{Solute transport and homogenization}
We obtain the effective dispersion by applying Mei's multiple-scale homogenization~\cite{mei96,mei10} to the Casson velocity. The method separates transverse diffusion, mean axial drift and longitudinal spreading through their different time scales. For a dilute neutral tracer, the starting equation is
\begin{equation}\label{eq:1003}
	\frac{\partial C^*}{ \partial t^*}+\nabla\cdot(\mathbf U C^*)=\nabla\cdot(D\nabla C^*).
\end{equation}

The tracer diffusivity $D>0$ is taken to be constant, including within the unyielded plug; it does not depend on the apparent viscosity. With the prescribed fully developed velocity and a cross-sectionally uniform instantaneous tracer release at $x^*=0$, Eq.~\eqref{eq:1003} reduces to

\begin{equation}\label{eq:17}
	\frac{\partial C^*}{ \partial t^*}+u^*\frac{\partial C^*}{\partial x^* }=D\left(\frac{\partial^2C^*}{\partial {x^*}^2}+\frac{\partial^2 C^*}{\partial {y^*}^2}\right)~~~~~~~-h<y^*<h,
\end{equation}
with the following initial and boundary conditions: 
\begin{equation}\label{eq:18}
	\left.C^*(x^*, y^*, t^*)\right|_{t^*=0}=C^0\delta\left(\frac{x^*}{h}\right),
\end{equation}
\begin{equation}\label{eq:19}
	\left[D\frac{\partial C^*}{\partial y^*}\right]_{y^*= h}=0,
\end{equation}
\begin{equation}\label{eq:19a}
	\left[D\frac{\partial C^*}{\partial y^*}\right]_{y^*= -h}=0,
\end{equation}
\begin{equation}	\label{eq:20}
	C^*(x^*,y^*,t^*)|_{x^*\rightarrow \pm\infty}=0.
\end{equation}

The delta distribution in Eq.~\eqref{eq:18} represents a longitudinal impulse applied uniformly across the channel section. Because $x^*/h$ is dimensionless, $C^0$ retains the units of concentration; the normalized initial condition has unit axial integral when expressed in $x^*/h$. Equations~\eqref{eq:19} and \eqref{eq:19a} impose impermeable walls, and Eq.~\eqref{eq:20} requires the plume to vanish far from the release location. This impulse normalization is distinct from the finite-width Gaussian release used for the direct numerical calculations.

The dimensionless time $t$ is defined by $t=Dt^*/h^2$, with $t^*$ the dimensional time. The stretched longitudinal variable used below is $x=x^*/L$. The separation between axial and transverse length scales introduces the small parameter

\begin{equation}
	\varepsilon=\frac{h}{L}(\ll 1).
\end{equation}

The three characteristic time scales follow from the channel dimensions and the chosen velocity scale. The electroosmotic normalization below requires $u_{HS}\neq0$; when that forcing vanishes, the nonzero reference velocity $U_0$ and $Pe_0$ in Eq.~\eqref{eq:yielding_pe} remain applicable.

The transverse diffusion time is

\begin{equation}
	T_0=O\left(\frac{h^2}{D}\right).
\end{equation}

The positive axial advection time is (for $u_{HS}\ne0$)

\begin{equation}
	T_1=O\left(\frac{L}{|u_{HS}|}\right).	
\end{equation}

The axial diffusion time is
\begin{equation}
	T_2=O\left(\frac{L^2}{D}\right).
\end{equation}
\\

The relative magnitudes of these time scales are
\begin{equation}\label{eq:24}
	T_0:T_1:T_2=1:\frac{1}{\varepsilon|Pe|}:\frac{1}{\varepsilon^2}.
\end{equation}

For the distinguished limit $|Pe|=O(1)$ as $\varepsilon\to0$, the transverse, advective and axial-diffusive time scales separate. The multiple time variables are introduced as
\begin{equation}\label{eq:25}
	t_0=t, t_1=\varepsilon t, t_2=\varepsilon^2t.
\end{equation}

The dimensionless variables are
\begin{equation}\label{eq:21}
	x=\frac{x^*}{L}, C=\frac{C^*}{C^0}, t=\frac{t^*}{h^2/D}.
\end{equation}

Using Eqs.~\eqref{eq:13} and \eqref{eq:21}, the transport equation~\eqref{eq:17} becomes
\begin{equation}\label{eq:22}
	\frac{\partial C}{ \partial t}+\varepsilon uPe\frac{\partial C}{\partial x}=\varepsilon^2\frac{\partial^2C}{\partial {x}^2}+\frac{\partial^2 C}{\partial {y}^2}~~~~~~~-1<y<1,
\end{equation}
where $Pe=u_{HS}h/D$ is the signed P\'eclet number based on the electroosmotic velocity scale; $|Pe|$ measures axial advection relative to molecular diffusion. The impermeable-wall conditions become

\begin{equation}\label{eq:23}
	\left[\frac{\partial C}{\partial y}\right]_{y= 1}=0,
\end{equation}
\begin{equation}\label{eq:23a}
	\left[\frac{\partial C}{\partial y}\right]_{y=-1}=0.
\end{equation}

The derivative with respect to time then separates into derivatives on the three scales,
\begin{equation}\label{eq:26}
	\frac{\partial}{\partial t}=\frac{\partial}{\partial t_0}+\varepsilon \frac{\partial}{\partial t_1}+\varepsilon^2\frac{\partial}{\partial t_2}.
\end{equation}

The concentration is sought as an asymptotic expansion
\begin{align}\label{eq:27}
	C(x,y,t) =C_0(x,y,t_0,t_1,t_2)+\varepsilon C_1(x,y,t_0,t_1,t_2)+\varepsilon^2 C_2(x,y,t_0,t_1,t_2)+\varepsilon^3 C_3(x,y,t_0,t_1,t_2)+O(\varepsilon^4).
\end{align}

Substitution of Eqs.~\eqref{eq:26} and \eqref{eq:27} into the transport equation~\eqref{eq:22} yields
\begin{align}\label{eq:28}
	&\left(\frac{\partial C_{0}}{\partial t_0}-\frac{\partial^2 C_{0}}{\partial y^2}\right)+\varepsilon\left(\frac{\partial C_{0}}{\partial t_1}+\frac{\partial C_{1}}{\partial t_0}+uPe\frac{\partial C_{0}}{\partial x}-\frac{\partial^2 C_{1}}{\partial y^2}\right)+\varepsilon^2\left(\frac{\partial C_{0}}{\partial t_2}+\frac{\partial C_{1}}{\partial t_1}\nonumber\right.&\\&\left.+\frac{\partial C_{2}}{\partial t_0}+uPe\frac{\partial C_{1}}{\partial x}-\frac{\partial^2 C_{0}}{\partial x^2}-\frac{\partial^2 C_{2}}{\partial y^2}\right)+\varepsilon^3\left(\frac{\partial C_{1}}{\partial t_2}+\frac{\partial C_{2}}{\partial t_1}+\frac{\partial C_{3}}{\partial t_0}\nonumber\right.&\\&\left.+uPe\frac{\partial C_{2}}{\partial x}-\frac{\partial^2 C_{1}}{\partial x^2}-\frac{\partial^2 C_{3}}{\partial y^2}\right)+O(\varepsilon^4)=0~~~~~~~~~~~~~~~ -1<y<1.&
\end{align}

The no-flux conditions give
\begin{equation}\label{eq:29}
	\frac{\partial C_{0}}{\partial y}+\varepsilon\,\frac{\partial C_{1}}{\partial y}+\varepsilon^2\,\frac{\partial C_{2}}{\partial y}+\varepsilon^3\,\frac{\partial C_{3}}{\partial y}+O(\varepsilon^4)=0 \mbox{ at }y=1,
\end{equation}
\begin{equation}\label{eq:29a}
	\frac{\partial C_{0}}{\partial y}+\varepsilon\,\frac{\partial C_{1}}{\partial y}+\varepsilon^2\,\frac{\partial C_{2}}{\partial y}+\varepsilon^3\,\frac{\partial C_{3}}{\partial y}+O(\varepsilon^4)=0 \mbox{ at }y=-1.
\end{equation}

Equating terms of equal order in $\varepsilon$ gives successive problems for the transverse concentration corrections. The expansion is developed through $O(\varepsilon^3)$; its accuracy at finite times is assessed separately from the long-time dispersion coefficient.

 \subsubsection{Leading-order problem}
At $O(\varepsilon^0)$, Eqs.~\eqref{eq:28} and \eqref{eq:29}--\eqref{eq:29a} give
\begin{equation}\label{eq:30}
	\frac{\partial C_{0}}{\partial t_0}=\frac{\partial^2 C_{0}}{\partial y^2},~~~~-1<y<1
\end{equation}
and
\begin{equation}\label{eq:31}
	\left.\frac{\partial C_{0}}{\partial y}\right|_{y=\pm 1}=0.
\end{equation}

The solution satisfying the homogeneous Neumann conditions is
\begin{equation}\label{eq:32}
	C_{0}=C_{0}^{(0)}(x,t_1,t_2) +\sum_{n=1}^{\infty} C_0^{(n)}(x,t_1,t_2)\cos\left(\frac{n\pi(y+1)}{2}\right) e^{-n^2\pi^2t_0/4}.
\end{equation}

Following decay of the transverse diffusion modes, Eq.~\eqref{eq:32} reduces to
\begin{equation}\label{eq:33}
	C_0=C_0(x,t_1,t_2).
\end{equation}

Thus, after relaxation of the transverse modes, $C_0$ is independent of $y$ and equals its own cross-sectional average $\langle C_0\rangle$. It represents the leading approximation to the cross-sectional mean of the full concentration; equality with $\langle C\rangle$ is asymptotic, not an exact identity for the finite-time transport equation.

 \subsubsection{First-order concentration correction}
At $O(\varepsilon)$, transverse uniformity of $C_0$ gives
\begin{equation}\label{eq:34}
	\frac{\partial C_{0}}{\partial t_1}+\frac{\partial C_{1}}{\partial t_0}+uPe\frac{\partial C_{0}}{\partial x}=\frac{\partial^2 C_{1}}{\partial y^2},
\end{equation}
and
\begin{equation}\label{eq:35}
	\left[\frac{\partial C_{1}}{\partial y}\right]_{y= \pm1}=0.
\end{equation}

For a cross-sectional quantity $f(y)$, the averaging operator is defined by
\begin{equation}\label{eq:37}
	\left<f\right>=\frac{1}{2}\int_{-1}^{1}f d y.
\end{equation}

The wall conditions and transverse uniformity of the leading mode imply $\langle C_0\rangle=C_0$ and $\left\langle\partial^2 C_n/\partial y^2\right\rangle=0$ for $n=1,2,3$.

The first-order cross-sectional solvability condition removes the mean advective contribution and leaves the velocity deviation that drives dispersion. Using the procedure of Mei \emph{et al.}~\cite{mei96,mei10} and the average~\eqref{eq:37}, we obtain (see Appendix~\ref{app:firstorder}, Eqs.~\eqref{eq:36}--\eqref{eq:39a})
\begin{equation}\label{eq:39}
	Pe\left(u-\langle u\rangle\right)\frac{\partial C_{0}}{\partial x}=\frac{\partial^2 C_{1}}{\partial y^2}.
\end{equation}

The first-order correction is written as
\begin{equation}\label{eq:40}
	C_1=Pe A_1(y)\frac{\partial C_0}{\partial x},
\end{equation}
where $A_1(y)$ is the first-order auxiliary function satisfying
\begin{equation}\label{eq:41}
	\frac{\dd^2 A_1}{\dd y^2}=u-\langle u\rangle~~~~~-1<y<1,
\end{equation}
with the boundary conditions 
\begin{equation}\label{eq:42}
	\left.\frac{\dd A_1}{\dd y}\right|_{y=\pm 1}=0,
\end{equation}
and 
\begin{equation}\label{eq:43}
	\left<A_1\right>=0.
\end{equation}

For the linear-viscosity velocity~\eqref{eq:16}, $A_1$ is obtained analytically as (with coefficients $d_1,\ldots,d_4$ defined in Appendix~\ref{app:coefficients})

\begin{equation*}
	A_1(y)=\frac{d_1-d_3}{2}y^2+\frac{d_2-d_1}{m^2\cosh m}\cosh my-\frac{d_2}{\kappa^2\cosh\kappa}\cosh \kappa y+d_4.
\end{equation*}

 \subsubsection{Second-order concentration correction}

At $O(\varepsilon^2)$, the transport equation and wall conditions yield
\begin{equation}\label{eq:44}
	\frac{\partial C_{0}}{\partial t_2}+\frac{\partial C_{1}}{\partial t_1}+\frac{\partial C_{2}}{\partial t_0}+uPe\frac{\partial C_{1}}{\partial x}=\frac{\partial^2 C_{0}}{\partial x^2}+\frac{\partial^2 C_{2}}{\partial y^2}~~~~~~~~-1<y<1,
\end{equation}
and 
\begin{equation}\label{eq:45}
	\left[\frac{\partial C_{2}}{\partial y}\right]_{y= \pm 1}=0.
\end{equation}

At the next order, the mean solvability condition supplies the longitudinal diffusion enhancement, while subtracting that mean yields the second transverse corrector. Following the homogenization steps in Refs.~\cite{mei96,mei10,da21,sa23} (Appendix~\ref{app:secondorder}, Eqs.~\eqref{eq:46}--\eqref{eq:53a}), the resulting equation is
\begin{align}\label{eq:53}
	\frac{\partial C_{2}}{\partial t_0}+	Pe^2\left[uA_1-\langle u\rangle A_1-\langle uA_1\rangle\right]\frac{\partial^2 C_{0}}{\partial x^2}=\frac{\partial^2 C_{2}}{\partial y^2}.	
\end{align}

We seek a correction of the form 
\begin{equation}\label{eq:54}
	C_2=Pe^2A_2(y)\frac{\partial^2 C_{0}}{\partial x^2}.
\end{equation}

Substitution into Eq.~\eqref{eq:53} gives the second-order auxiliary boundary-value problem

\begin{equation}\label{eq:55}
	\frac{\dd^2 A_2}{\dd y^2}=uA_1-\langle u\rangle A_1-\langle uA_1\rangle,
\end{equation}
\begin{equation}\label{eq:56}
	\left.\frac{\dd A_2}{\dd y}\right|_{y=\pm 1}=0,
\end{equation}
and
\begin{equation}\label{eq:57}
	\left<A_2\right>=0.
\end{equation}

The auxiliary problem~\eqref{eq:55}--\eqref{eq:57}, with homogeneous Neumann conditions and zero mean, uniquely determines the higher-order auxiliary function $A_2$. It can be evaluated without lengthy hyperbolic expansions by the exact Green-operator representation in Appendix~\ref{app:neumann_operator}. This representation remains valid for both linear-viscosity and Casson velocities.

\subsubsection{Third-order concentration correction}
At $O(\varepsilon^3)$, the corresponding equations are
\begin{equation}\label{eq:58}
	\frac{\partial C_{1}}{\partial t_2}+\frac{\partial {C_{2}}}{\partial t_1}+\frac{\partial C_3}{\partial t_0}+uPe{\frac{\partial C_{2}}{\partial x}}=\frac{\partial^2 C_{1}}{\partial x^2}+\frac{\partial^2 {C_{3}}}{\partial y^2},
\end{equation}
and 
\begin{equation}\label{eq:59}
	\left[\frac{\partial C_{3}}{\partial y}\right]_{y=\pm1}=0.
\end{equation}

Averaging and subtracting the mean as in Refs.~\cite{ba19,da21} gives (see Appendix~\ref{app:thirdorder})
\begin{align}\label{eq:70}
	\frac{\partial C_{3}}{\partial t_0}+	Pe^3\left[uA_2-\langle u\rangle A_2-\left<uA_2\right>-\langle uA_1\rangle A_1\right]\frac{\partial^3 C_0}{\partial x^3}=\frac{\partial^2 C_3}{\partial y^2}.
\end{align}

We write the third-order correction as
\begin{equation}\label{eq:71}
	C_3=Pe^3A_3(y)\frac{\partial^3 C_0}{\partial x^3},
\end{equation}
where the third-order auxiliary function $A_3$ satisfies the following boundary-value problem:
\begin{equation}\label{eq:72}
	\frac{\dd^2 A_3}{\dd y^2}=uA_2-\langle u\rangle A_2-\left<uA_2\right>-\langle uA_1\rangle A_1,
\end{equation}
\begin{equation}\label{eq:73}
	\left.\frac{\dd A_3}{\dd y}\right|_{y=\pm 1}=0,
\end{equation}
\begin{equation}\label{eq:74}
	\left<A_3\right>=0.
\end{equation}

The third-order auxiliary function is therefore determined by its boundary-value problem~\eqref{eq:72}--\eqref{eq:74} or equivalently by the zero-mean Neumann integral representation~\eqref{eq:green_higher_auxiliary}; the intermediate solvability steps are given in Appendix~\ref{app:thirdorder}, Eqs.~\eqref{eq:60}--\eqref{eq:70a}.

 \subsubsection{Taylor dispersion coefficient}

The successive solvability conditions now close the cross-sectionally averaged transport equation: the first-order condition determines the mean drift~\eqref{eq:38}, and the second-order condition determines the effective axial diffusivity~\eqref{eq:75}. Together they give (Appendix~\ref{app:homogenization})
\begin{align}\label{eq:77}
	\frac{\partial C_0}{\partial t}+\varepsilon Pe\langle u\rangle\frac{\partial C_0}{\partial x}=\varepsilon^2\left[1-Pe^2\langle uA_1\rangle\right]\frac{\partial^2 C_0}{\partial x^2}.
\end{align}

In the moving coordinate $\xi=x/\varepsilon-Pe\langle u\rangle t$ and time $\tau=t$, Eq.~\eqref{eq:77} becomes
\begin{equation}\label{eq:78}
	\frac{\partial C_0}{\partial \tau}=D_T\frac{\partial^2 C_0}{\partial \xi^2},
\end{equation}
where
\begin{equation}\label{eq:79}
D_T=\frac{D_{\mathrm{eff}}}{D}=1+Pe^2D_T^*,\qquad
D_T^*=-\langle uA_1\rangle
=\left\langle\left(\frac{\dd A_1}{\dd y}\right)^2\right\rangle.
\end{equation}
Here $D_{\mathrm{eff}}$ is the dimensional long-time effective axial diffusivity, $D_T$ its dimensionless value including molecular diffusion, and $D_T^*$ the nonnegative shear-induced dispersion factor for the linear-viscosity profile. Equation~\eqref{eq:yielding_dispersion} defines the corresponding normalized Casson quantity $D_T^{\mathrm C}$. Both are total normalized diffusivities; their values differ when the underlying constitutive models produce different flow profiles. The factor $D_T^*$ is not an additional diffusivity and is not a time-dependent dispersion coefficient.

For the idealized impulse release, the leading-order homogenized concentration is
\begin{equation}\label{eq:80}
	C_0=\frac{1}{\sqrt{4\pi D_T\tau}}\exp\left(-\frac{\xi^2}{4D_T\tau}\right),	
\end{equation} 

The linear-viscosity dispersion factor $D_T^*$ can be evaluated from $A_1$ using the energy identity in Eq.~\eqref{eq:positive_dispersion}. Its exact evaluation in elementary hyperbolic functions is given in Appendix~\ref{app:coefficients}; this formulation avoids unnecessary cancellation-prone higher-order coefficients. For the same prescribed velocity profile, the corresponding Casson and linear-model normalizations are equivalent by Eq.~\eqref{eq:reference_yielding_conversion}; the constitutive laws themselves remain distinct.

Multiplication of Eq.~\eqref{eq:41} by $A_1$ and integration across the channel give a positive-definite expression for the dispersive enhancement. The same energy identity applies to the nonlinear Casson velocity through Eq.~\eqref{eq:yielding_dispersion}:
\begin{equation}\label{eq:positive_dispersion}
D_T=1-Pe^2\langle uA_1\rangle
=1+Pe^2\left\langle\left(\frac{\dd A_1}{\dd y}\right)^2\right\rangle\geq1.
\end{equation}
Equation~\eqref{eq:positive_dispersion} relates the additional axial diffusivity to the cross-sectional velocity variation and the corresponding Taylor--Aris auxiliary function. In the pressure-driven Casson limit~\eqref{eq:casson_exact}, and using $U_P$ as the velocity scale, the result near $\delta=1-\alpha=0$ is
\begin{equation}\label{eq:near_disp}
\frac{D_{\mathrm{eff}}}{D}-1
=\left(\frac{U_Ph}{D}\right)^2\frac{\delta^8}{6912}
+O\left(\left(\frac{U_Ph}{D}\right)^2\delta^9\right).
\end{equation}
The cubic decay of the mean velocity~\eqref{eq:nearflux} and eighth-power decay of the dispersion enhancement~\eqref{eq:near_disp} characterize approach to cessation of pressure-driven Casson flow at a fixed imposed pressure gradient. Their derivation, given in Appendix~\ref{app:near_arrest}, assumes a constant molecular diffusivity and the plane-channel geometry.

\subsection{Numerical methods}\label{sec:numerical_methods}
The nonlinear Casson stress and velocity equations~\eqref{eq:nonlinear_stress}--\eqref{eq:nonlinear_shear} are solved on $0\leq y\leq1$ by shooting~\cite{AscherMattheijRussell1995}. The centreline velocity is adjusted to satisfy no slip, and the yield fronts are located from $|\widehat\tau|=Y$. The computed velocity then enters the auxiliary problem~\eqref{eq:yielding_auxiliary} derived by homogenization; quadrature of Eq.~\eqref{eq:yielding_dispersion} gives $D_T^{\mathrm C}$. We locate magnetic dispersion maxima by parameter scanning and bounded refinement. For fixed-throughput calculations, $\Pi$ is adjusted to impose the prescribed mean velocity.

For finite-time transport, the advection--diffusion equation~\eqref{eq:17} is solved in the moving coordinate~\eqref{eq:yielding_pe} using spectral discretization in the axial direction~\cite{Trefethen2000}, conservative finite-volume transverse diffusion~\cite{EymardGallouetHerbin2000}, and Strang operator splitting~\cite{Strang1968}. The wall-normal flux vanishes, and the axial periodic interval is sufficiently long to avoid interaction of the concentration tails with its boundaries over the simulated interval. An independently solved finite-volume moment hierarchy~\eqref{eq:direct_moments_new} checks the axial variance. Analytical benchmarks, spatial-grid refinement, and time-step comparisons are reported in Secs.~\ref{sec:casson_numeric_validation_new} and \ref{sec:direct_2d_casson_new}. A two-component Casson calculation separately examines sensitivity to transverse shear in Sec.~\ref{sec:transverse_sensitivity}.

\section{Results and discussion}

Analytical limits and published electroosmotic and electromagnetic results are used first to check the velocity and dispersion calculations. We then apply the homogenized diffusivity to the Casson flow, tracing the effects of yield surfaces, magnetic forcing and operating constraints. Direct transport calculations provide an independent account of the evolving concentration field.

\subsection{Validation against published results}
\begin{figure}
		\centering
\includegraphics[width=0.90\linewidth]{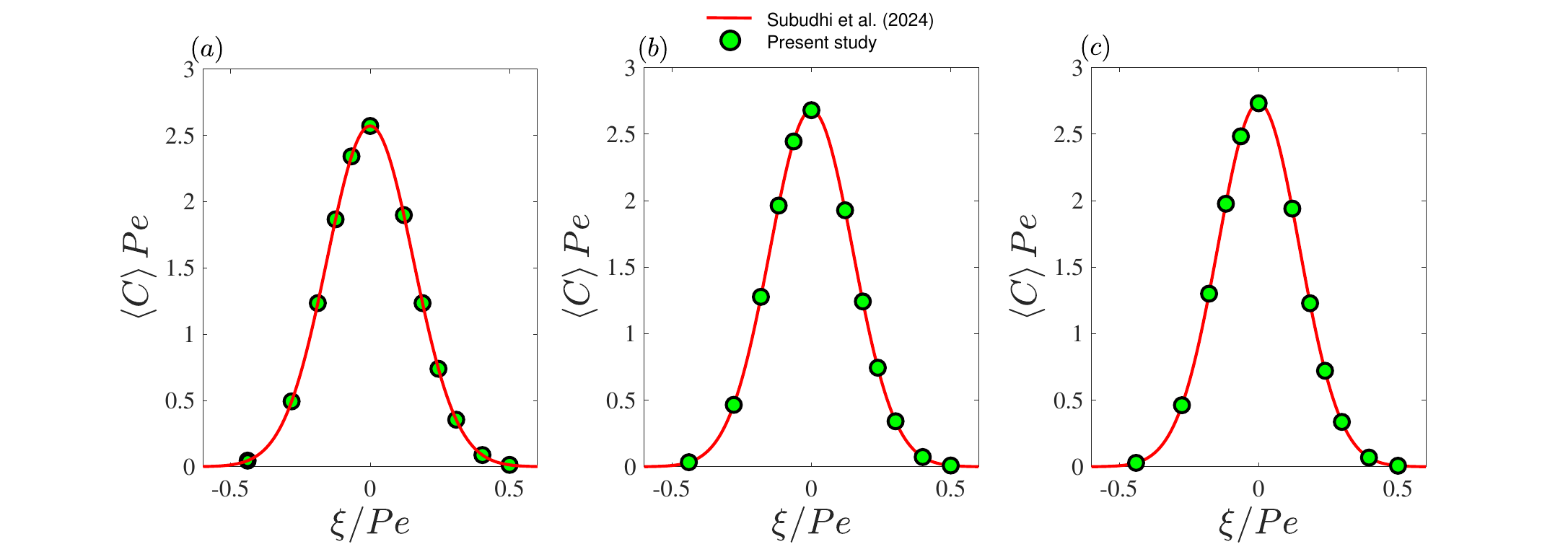}
	\caption{Mean concentration profiles for $\kappa=$ (a) 10, (b) 15, and (c) 20, compared with Subudhi \emph{et al.} (2024)~\cite{Subudhi24}. Here $Ha=0$, $\beta\to\infty$, $Pe=10$, and the imposed pressure gradient vanishes.}
	\label{fig:match}
\end{figure}

\begin{figure}
		\centering
\includegraphics[width=0.60\linewidth]{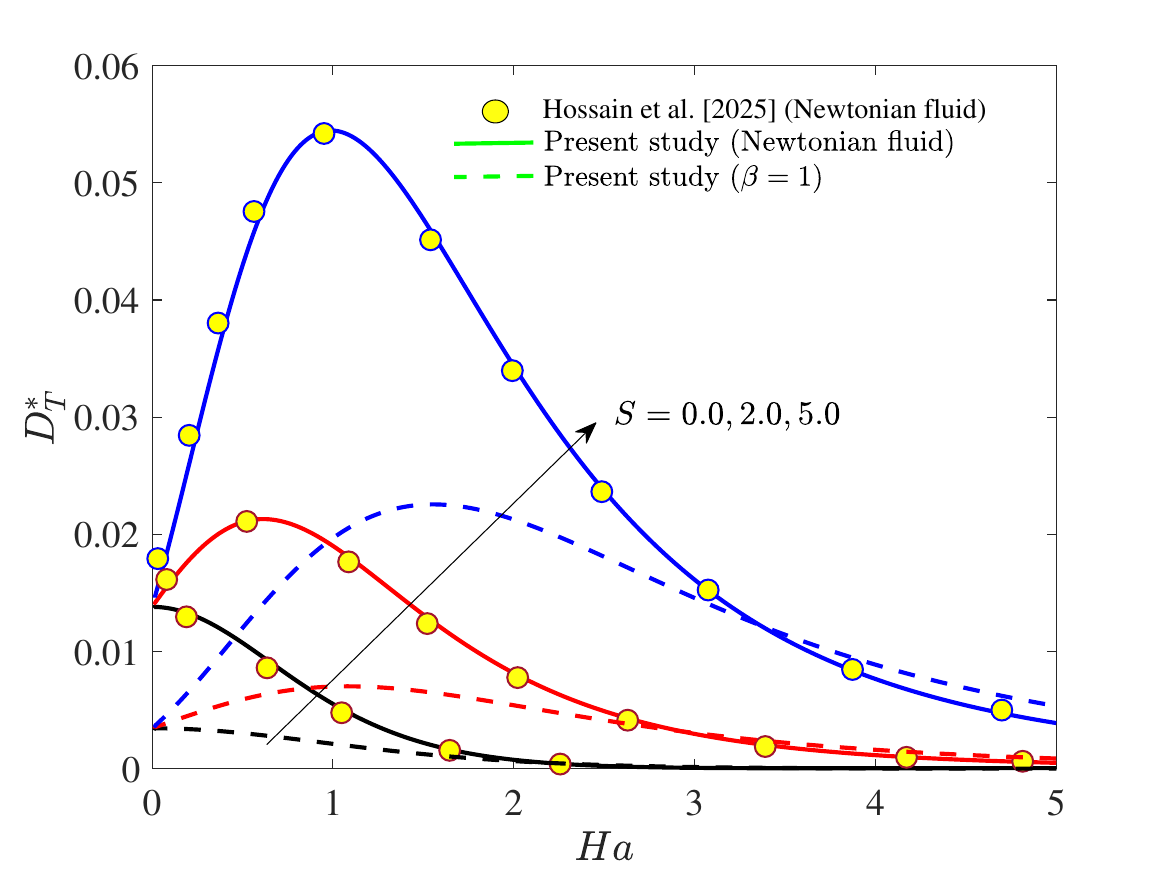}
	\caption{Comparison of $D_T^{*}$ versus $Ha$ for $S=0,2,5$, comparing the present results with Hossain \emph{et al.} (2025)~\cite{Hossain2025}. Parameters: $\kappa=4$ and $\beta_1=1$.}
	\label{fig:match3}
\end{figure}

\begin{table}
\centering
\renewcommand{\arraystretch}{1.2}
\begin{tabular}{lcccccc}
\hline
\textbf{Model} & \multicolumn{6}{c}{$S$} \\
\cline{2-7}
 & 0.5 & 1 & 2 & 3 & 4 & 5 \\
\hline
Hossain \textit{et al.} (2025)~\cite{Hossain2025} (Newtonian)
& 3.9460 & 7.9688 & 16.2445 & 24.8271 & 33.7167 & 42.9132 \\

Present ($\beta\rightarrow\infty$)
& 3.9460 & 7.9688 & 16.2445 & 24.8271 & 33.7167 & 42.9132 \\

Present ($\beta=10$)
& 3.9457 & 7.9681 & 16.2432 & 24.8251 & 33.7139 & 42.9096 \\

Present ($\beta=1$)
& 3.9443 & 7.9653 & 16.2372 & 24.8159 & 33.7012 & 42.8931 \\
\hline
\end{tabular}
\caption{Percentage change $I$ in dispersion with the spanwise electric-field parameter $S$. Newtonian values from Hossain \emph{et al.} (2025)~\cite{Hossain2025} are compared with the calculated linear-viscosity results for different $\beta$.}
\label{tab:comparison}
\end{table}

Figure~\ref{fig:match} compares the calculated mean concentration profiles with the Newtonian electroosmotic solution of Subudhi \emph{et al.}~\cite{Subudhi24} for $\kappa=10$, $15$ and $20$. The Newtonian limit is obtained by taking $\beta\to\infty$ and $Ha=0$. Agreement for all three EDL thicknesses provides a benchmark for the electroosmotic concentration calculation.

Figure~\ref{fig:match3} compares $D_T^*$ against $Ha$ with the Newtonian calculations of Hossain \emph{et al.}~\cite{Hossain2025} at $S=0$, $2$ and $5$. The curves agree in the Newtonian limit, while finite $\beta$ increases the effective viscosity and reduces dispersion for the illustrated forcing. This constitutive comparison involves no yield surface; the stress-dependent Casson response is considered separately below. Table~\ref{tab:comparison} quantifies the relative change in $D_T^*$ with $S$, defined by
\begin{equation}
I=\frac{\left(D_T^{*}\right)_{S\neq0}-\left(D_T^{*}\right)_{S=0}}
{\left(D_T^{*}\right)_{S=0}}\times100\%.
\end{equation}
In Table~\ref{tab:comparison}, the $\beta\to\infty$ values match the published Newtonian increments, while finite $\beta$ produces a modest reduction.
\subsection{Pressure-driven dispersion near arrest}
The pressure-driven Casson solution~\eqref{eq:casson_exact} gives the yield-surface position directly. Casson-flow dispersion in channels and tubes was examined by Dash \emph{et al.}~\cite{DashJayaramanMehta2000}; the present limiting analysis is used to establish the numerical benchmark and the asymptotic scaling under a fixed imposed pressure gradient. For $0<\alpha<1$, the central plug occupies $|y|\leq\alpha$, bounded by two sheared layers; for $\alpha\geq1$, no flow is possible under no-slip conditions. Equation~\eqref{eq:casson_mean} recovers the Newtonian mean velocity $U_P/3$ as $\alpha\to0$ and approaches zero as $\delta^3/12$ when $\delta=1-\alpha\to0^+$. These limits provide stringent checks on the nonlinear stress--velocity solver.

\begin{figure}
\centering
\includegraphics[width=0.5\linewidth]{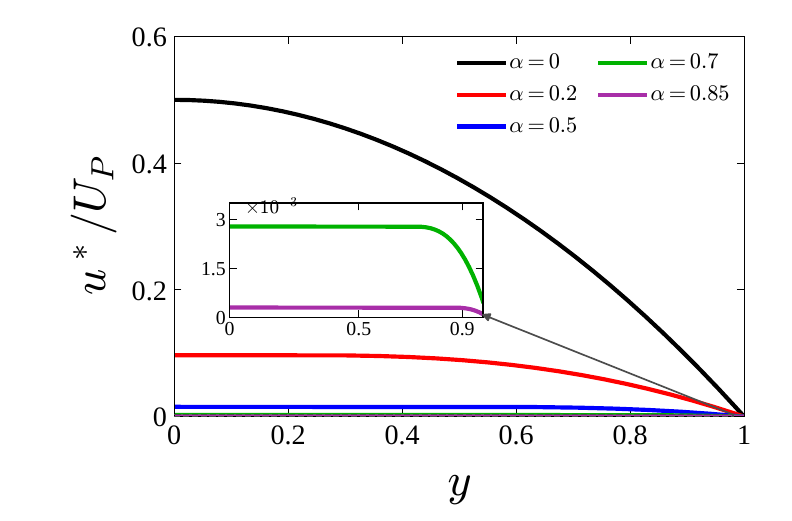}
\caption{Exact pressure-driven Casson velocity profiles for selected yield-to-wall-stress ratios $\alpha$. The inset magnifies the low-velocity region for $\alpha=0.7$ and $0.85$.}
\label{fig:exact_casson_velocity_new}
\end{figure}

The velocity profiles in Fig.~\ref{fig:exact_casson_velocity_new} show a progressively wider unyielded core and weaker flow as $\alpha$ approaches unity. The inset resolves the small but nonzero velocities for $\alpha=0.7$ and $0.85$.

Near arrest, the narrow yielded layers reduce both the throughput and the cumulative velocity departure from its mean. Substitution of the exact Casson profile into the Taylor--Aris auxiliary problem~\eqref{eq:41} gives the eighth-power dispersion law~\eqref{eq:near_disp}, whereas the mean flow decays cubically~\eqref{eq:nearflux}. Thus, shear-induced spreading becomes negligible faster than the mean transport as the yield stress approaches the wall stress. This pressure-driven limit has a known stress distribution; the asymptotic coefficients are derived in Appendix~\ref{app:near_arrest}.

\begin{figure*}
\centering
\begin{subfigure}{0.49\linewidth}
\centering
\includegraphics[width=\linewidth]{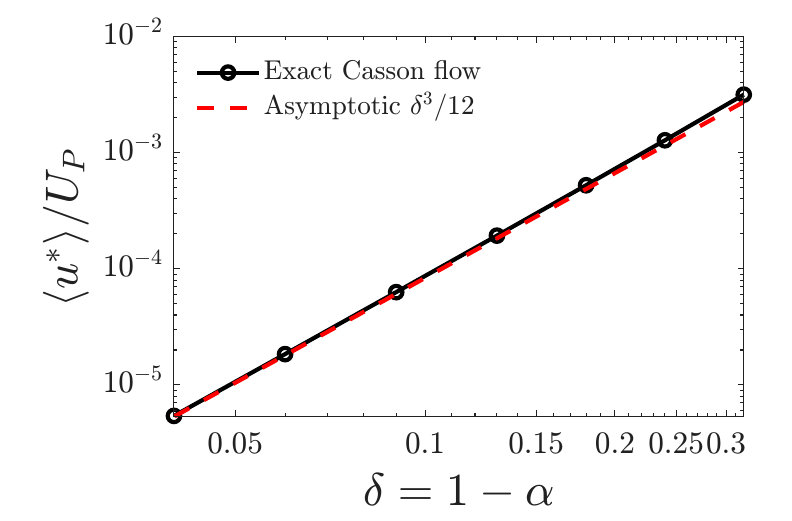}
\caption{}\label{fig:near_arrest_flux_new}
\end{subfigure}\hfill
\begin{subfigure}{0.49\linewidth}
\centering
\includegraphics[width=\linewidth]{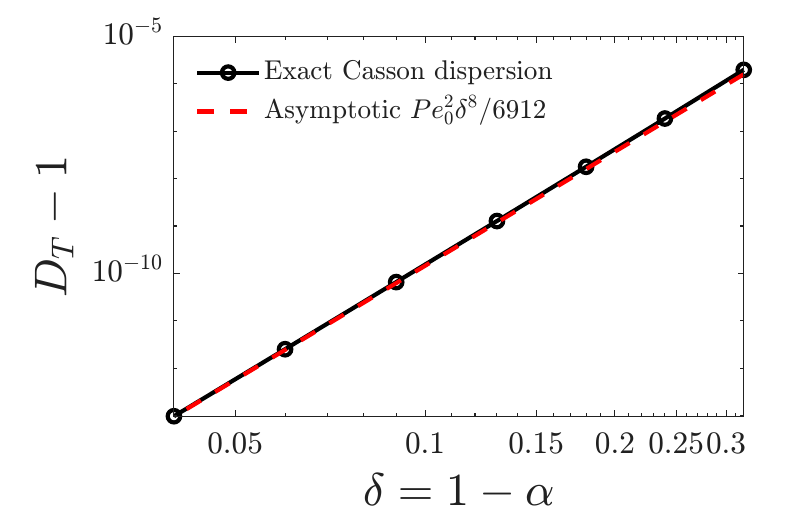}
\caption{}\label{fig:near_arrest_disp_new}
\end{subfigure}
\caption{Exact pressure-driven Casson results in the near-arrest limit: (a) normalized mean axial velocity and the cubic asymptotic law; (b) shear-induced Taylor-dispersion enhancement and the eighth-power asymptotic law. Here $\delta=1-\alpha$, and $U_0=U_P$ so that $Pe_0=U_Ph/D=10$; the externally imposed pressure gradient is held fixed. The applied magnetic and electric fields vanish in this limit.}
\label{fig:near_arrest_new}
\end{figure*}

\subsection{Dispersion in the linear-viscosity model}

\begin{figure*}
	\centering
	\begin{subfigure}{0.47\linewidth}
		\centering
		\includegraphics[width=\linewidth]{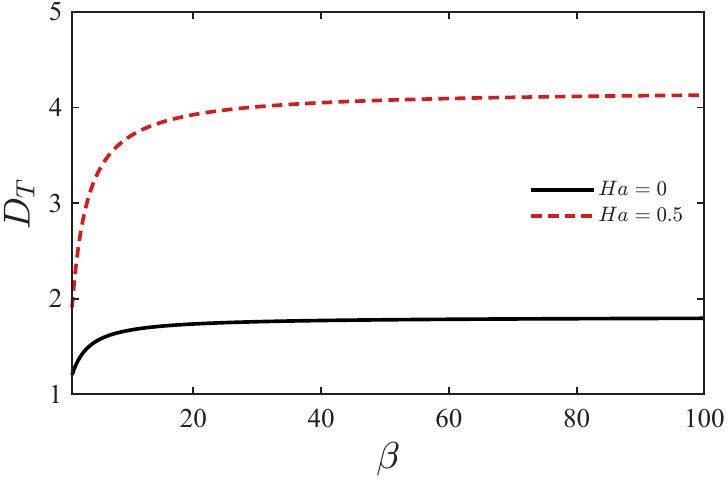}
		\caption{}
		\label{fig:2a}
	\end{subfigure}
	\begin{subfigure}{0.47\linewidth}
		\centering
		\includegraphics[width=\linewidth]{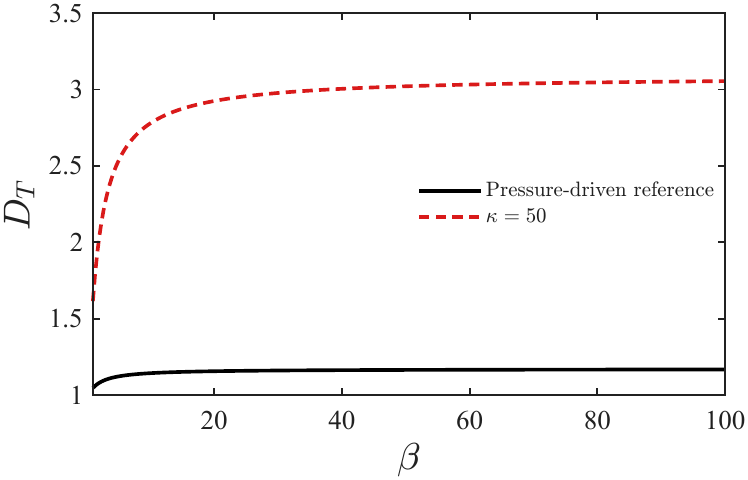}
		\caption{}
		\label{fig:2b}
	\end{subfigure}
	\begin{subfigure}{0.47\linewidth}
		\centering
		\includegraphics[width=\linewidth]{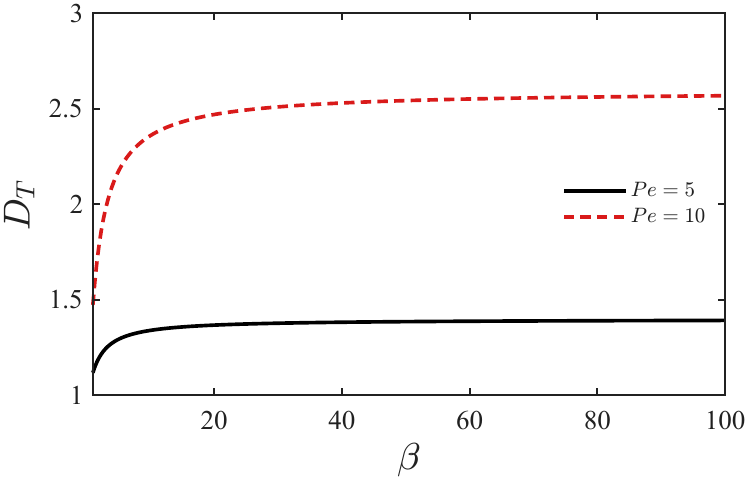}
		\caption{}
		\label{fig:2c}
	\end{subfigure}
	\begin{subfigure}{0.47\linewidth}
		\centering
		\includegraphics[width=\linewidth]{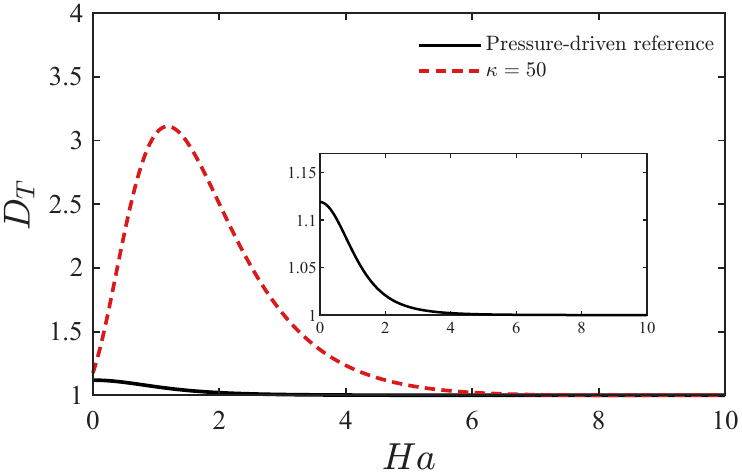}
		\caption{}
		\label{fig:2d}
	\end{subfigure}
	\begin{subfigure}{0.47\linewidth}
		\centering
		\includegraphics[width=\linewidth]{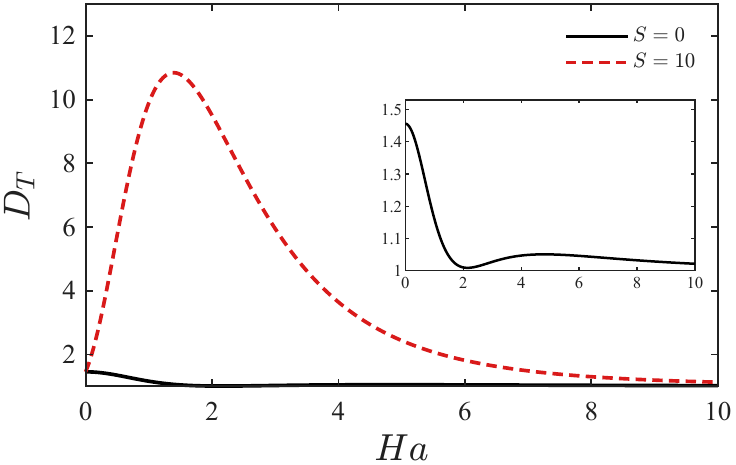}
		\caption{}
		\label{fig:2e}
	\end{subfigure}
	\begin{subfigure}{0.47\linewidth}
		\centering
		\includegraphics[width=\linewidth]{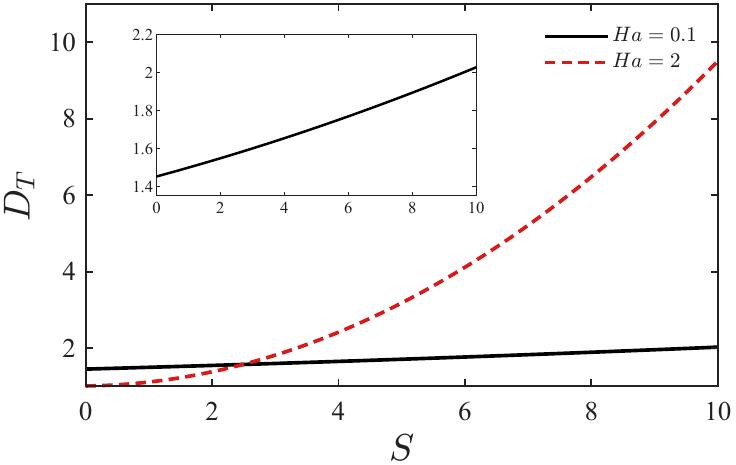}
		\caption{}
		\label{fig:2f}
	\end{subfigure}
	\caption{Normalized Taylor--Aris diffusivity $D_T$ for the linear-viscosity model: (a) variation with $\beta$ at $Ha=0$ and $0.5$ ($\kappa=10$, $S=5$); (b) variation with $\beta$ for $\kappa=50$ and the pressure-driven magnetic reference ($Ha=0.5$); (c) variation with $\beta$ at $Pe=5$ and $10$ ($Ha=0.5$, $\kappa=40$, $S=4$); (d) dependence on $Ha$ for $\kappa=50$ and the pressure-driven reference ($\beta=3$, $S=5$ for the electrically forced curve); (e) dependence on $Ha$ at $S=0$ and $10$ ($\beta=3$, $\kappa=10$); and (f) variation with $S$ at $Ha=0.1$ and $2$ ($\beta=3$, $\kappa=10$). The electrically forced cases use $\beta_1=1$ and $Pe=10$ unless $Pe$ is varied. In panels (b) and (d), the pressure-driven reference is computed separately with $E_x=E_z=0$, $\Pi_{\mathrm{ref}}=1$ and $Pe_{\mathrm{ref}}=10$, retaining magnetic braking under the stipulated current closure. Its normalization is given by Eqs.~\eqref{eq:pressure_only_scaling}--\eqref{eq:pressure_only_exact}.}
	\label{fig:2}
\end{figure*}
For the linear-viscosity law~\eqref{eq:9}, the dispersion coefficient depends on $\beta$, $Ha$, $\kappa$, $S$ and $Pe$ through the analytical velocity~\eqref{eq:16}, as illustrated in Fig.~\ref{fig:2}. An increase in $\beta$ decreases $\mu_{\mathrm{eff}}$ and can strengthen the velocity variation responsible for Taylor dispersion. At a fixed velocity profile, the dispersive enhancement scales as $Pe^2$ according to Eq.~\eqref{eq:positive_dispersion}. The dependence on $\kappa$ reflects the change in electroosmotic forcing across the EDL and the corresponding shear distribution.

The spanwise-current contribution to the axial momentum equation scales with $S Ha$, whereas magnetic braking scales with $Ha^2u$. As $Ha$ increases, the competition initially strengthens and then weakens the velocity variation for the cases in Fig.~\ref{fig:2}, producing a dispersion maximum. A related Newtonian response was reported by Hossain \emph{et al.}~\cite{Hossain2025}. The peak location depends on the imposed electrical forcing and the effective viscosity.

The relationship between electrolyte properties and the independently varied dimensionless parameters is examined in Sec.~\ref{sec:electromagnetic_constraints}.

\subsection{Magnetic dispersion extrema in the linear-viscosity model}

\begin{figure}
	\centering
	\begin{subfigure}{0.47\linewidth}
		\centering
		\includegraphics[width=\linewidth]{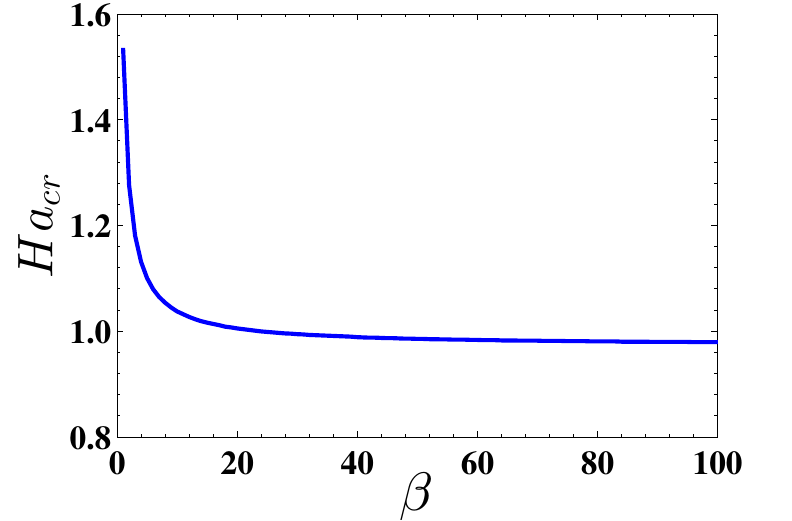}
		\caption{}
		\label{fig:3a}
	\end{subfigure}
	\begin{subfigure}{0.47\linewidth}
		\centering
		\includegraphics[width=\linewidth]{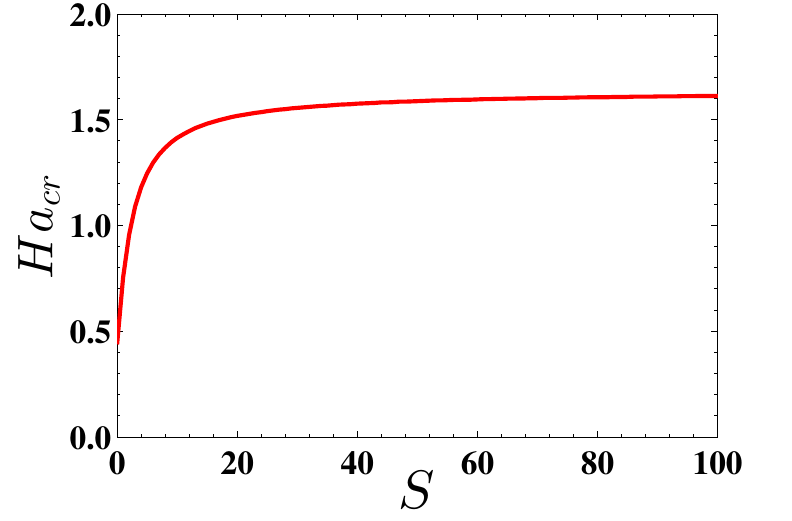}
		\caption{}
		\label{fig:3b}
	\end{subfigure}
	\caption{Hartmann number $Ha_{\mathrm{cr}}$ at maximum dispersion for the linear-viscosity law versus (a) viscosity parameter $\beta$ and (b) spanwise electric-field parameter $S$. Unless varied, $\kappa=10$, $\beta_1=1$, $S=5$, $\beta=1$, and $Pe=10$.}
	\label{fig:3}
\end{figure}

The Hartmann number $Ha_{\mathrm{cr}}$ is defined by a local extremum of the dispersion curve, $\partial D_T/\partial Ha=0$, with the remaining parameters fixed. Figure~\ref{fig:3} shows its variation with $\beta$ and $S$ under the linear-viscosity law. Increasing $S$ strengthens the current-driven Lorentz contribution and displaces the peak toward larger $Ha$; varying $\beta$ changes the competing viscous and magnetic terms. The nonlinear Casson results below determine the corresponding magnetic maximum together with the yield-surface structure.

\subsection{Yield surfaces and magnetic-dispersion optima}\label{sec:actual_casson_emhd}

We next resolve the yield surfaces and dispersion of the Casson fluid under combined pressure and electromagnetic forcing. Unless stated otherwise, we take $\mu=\mu_p$, $\Pi=1$, $\Omega=0.12$, $S_p=1$, $\kappa=10$ and $Pe_0=10$ with no slip at the channel walls. As $Ha_p$ and $Y$ vary, $\Pi$, $\Omega$ and $S_p$ remain fixed within the dimensionless axial model, while the mean throughput adjusts in response to the flow field. The parameter sweeps treat $Ha_p$ and $\kappa$ as independent dimensionless parameters; as explained in Sec.~\ref{sec:electromagnetic_constraints}, this is a theoretical sweep and does not correspond automatically to varying the magnetic field at fixed electrolyte properties. This setting isolates the coupled effects of yielding and magnetic forcing on the dispersion maximum.

Using the shooting procedure described in Sec.~\ref{sec:numerical_methods}, we determine the coupled stress and velocity for each parameter set without prescribing the yield-front position. The calculated velocity determines the normalized dispersion coefficient through Eqs.~\eqref{eq:yielding_auxiliary} and \eqref{eq:yielding_dispersion}.

\begin{figure*}
\centering
\begin{subfigure}{0.49\linewidth}
\centering
\includegraphics[width=\linewidth]{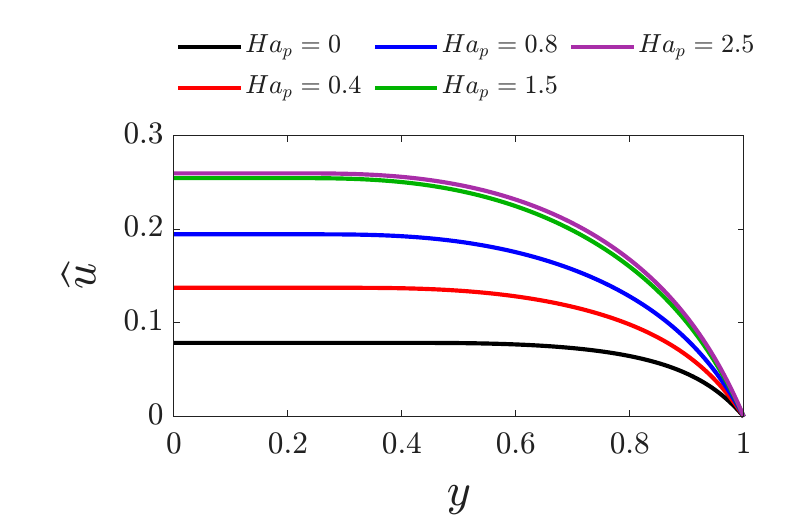}
\caption{}\label{fig:true_casson_velocity_new}
\end{subfigure}\hfill
\begin{subfigure}{0.49\linewidth}
\centering
\includegraphics[width=\linewidth]{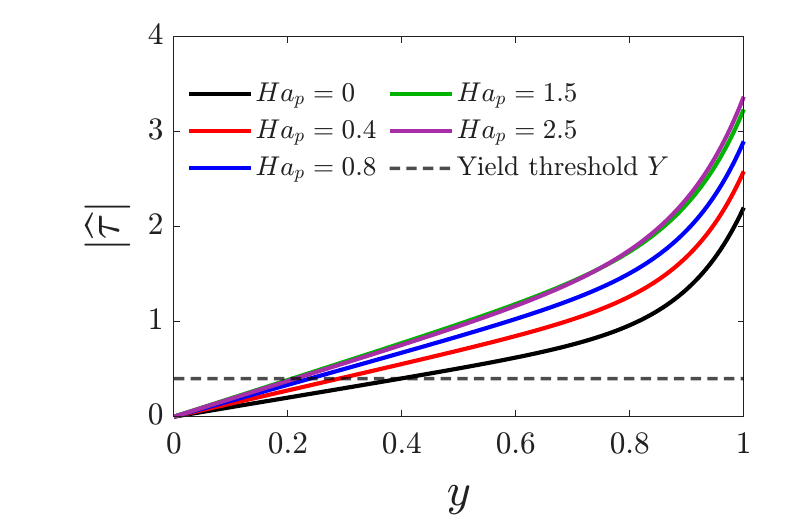}
\caption{}\label{fig:true_casson_stress_new}
\end{subfigure}
\caption{Stress-dependent Casson--EMHD profiles versus $Ha_p$: (a) axial velocity and (b) shear-stress magnitude and yield threshold. Here $\Pi=1$, $\Omega=0.12$, $S_p=1$, $\kappa=10$, and $Y=0.4$.}
\label{fig:true_casson_flow_new}
\end{figure*}

\begin{figure*}[t]
\centering
\begin{subfigure}[t]{0.49\linewidth}
\centering
\includegraphics[width=\linewidth]{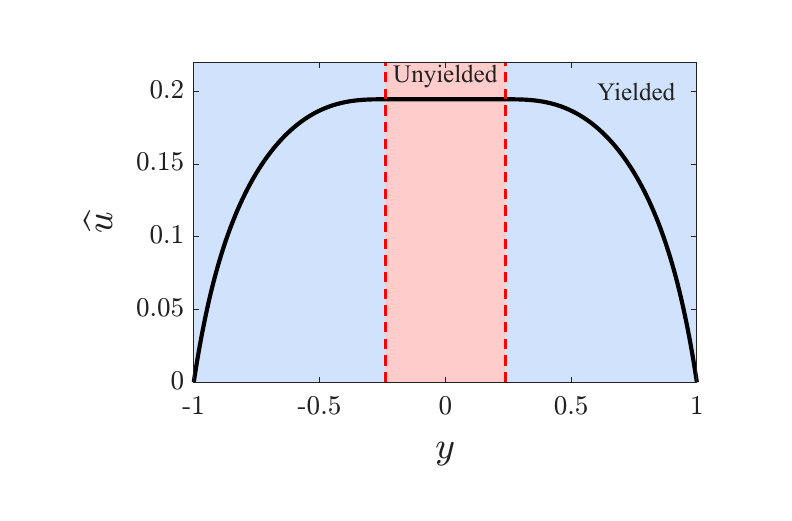}
\caption{}\label{fig:yield_region_y04_new}
\end{subfigure}\hfill
\begin{subfigure}[t]{0.49\linewidth}
\centering
\includegraphics[width=\linewidth]{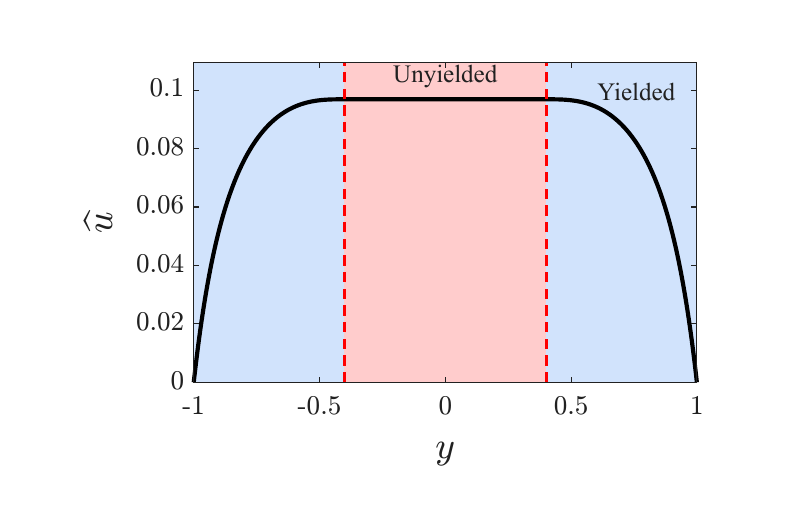}
\caption{}\label{fig:yield_region_y07_new}
\end{subfigure}
\caption{Yielded and unyielded regions of Casson flow at (a) $Y=0.4$ and (b) $Y=0.7$, with $Ha_p=0.8$, $\Pi=1$, $\Omega=0.12$, $S_p=1$, and $\kappa=10$. Dashed lines mark $y=\pm y_f$; shaded central regions are unyielded.}
\label{fig:yielded_unyielded_visual_new}
\end{figure*}

Figures~\ref{fig:true_casson_flow_new} and \ref{fig:yielded_unyielded_visual_new} show how magnetic forcing changes the stress distribution, the sheared-layer velocity and the size of the unyielded core. At $Ha_p=0.8$, the yield front lies at $|y|\simeq0.238$ for $Y=0.4$ and at $|y|\simeq0.401$ for $Y=0.7$. These positions characterize the extent of yielding but do not, on their own, determine the dispersion coefficient. Integrating Eq.~\eqref{eq:yielding_auxiliary} from the centreline gives the cumulative velocity-deviation function
\begin{equation}\label{eq:casson_cumulative_variation_new}
\mathcal{Q}(y)=\int_0^y\bigl(\widehat u(s)-\langle\widehat u\rangle\bigr)\,\dd s,
\qquad
D_T^{\mathrm C}-1=Pe_0^2\int_0^1\mathcal{Q}(y)^2\,\dd y.
\end{equation}
Because the plug velocity generally differs from the cross-sectional mean, an unyielded region can contribute to the cumulative deviation $\mathcal Q$ even though its local shear rate is zero. Consequently, the plug width alone does not determine dispersion; the entire velocity distribution and the operating constraint enter Eq.~\eqref{eq:casson_cumulative_variation_new}.

Magnetic-dispersion maxima are identified by scanning $0\leq Ha_p\leq4$ and refining the interior local maxima of $D_T^{\mathrm C}(Ha_p)$ through bounded optimization. Each maximizing state is recalculated on a refined transverse grid. The values in Table~\ref{tab:true_casson_peaks_new} are therefore resolved optima within the stated magnetic interval at the prescribed pressure and electric forcing.

\begin{table*}
\caption{Numerically refined magnetic-dispersion maxima of the stress-dependent Casson model for $\Pi=1$, $\Omega=0.12$, $S_p=1$, $\kappa=10$ and $Pe_0=10$. The unyielded fraction is the proportion of the slit with $|\widehat\tau|\leq Y$; for a symmetric central plug it equals the half-channel yield-front position $y_f$. The slight differences in tabulated last digits reflect independent grid-counting and interpolated-front estimates. A dash denotes absence of an internal yield front in the Newtonian limit $Y=0$.}
\label{tab:true_casson_peaks_new}
\centering
\begin{tabular}{ccccc}
\toprule
$Y$ & $Ha_{p,\mathrm{opt}}$ & $D_{T,\mathrm{max}}^{\mathrm C}$ & Unyielded fraction & First yield front $y_f$ \\
\midrule
$0$   & $0.69554$ & $1.45357856$ & $0$ & --- \\
$0.2$ & $1.43661$ & $1.07480367$ & $0.11854$ & $0.11853$ \\
$0.4$ & $1.93346$ & $1.03413003$ & $0.20563$ & $0.20558$ \\
$0.7$ & $2.63351$ & $1.01468659$ & $0.30271$ & $0.30274$ \\
\bottomrule
\end{tabular}
\end{table*}

\begin{figure*}
\centering
\begin{subfigure}{0.49\linewidth}
\centering
\includegraphics[width=\linewidth]{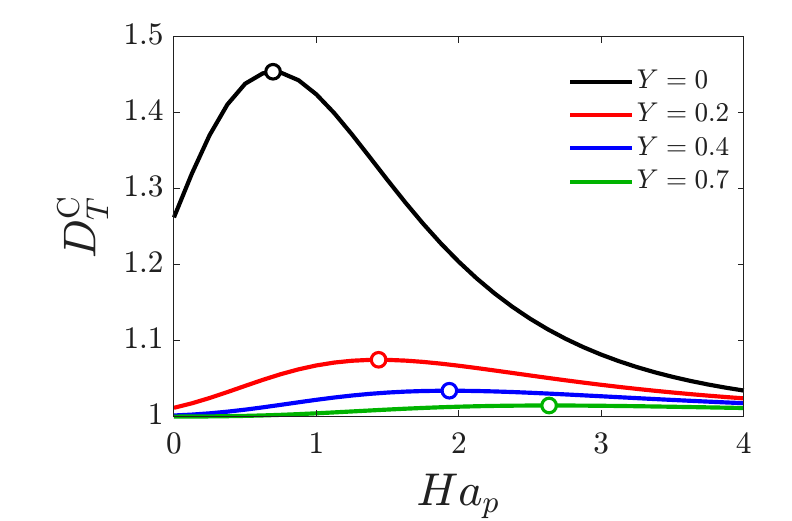}
\caption{}\label{fig:actual_peak_curves_new}
\end{subfigure}\hfill
\begin{subfigure}{0.49\linewidth}
\centering
\includegraphics[width=\linewidth]{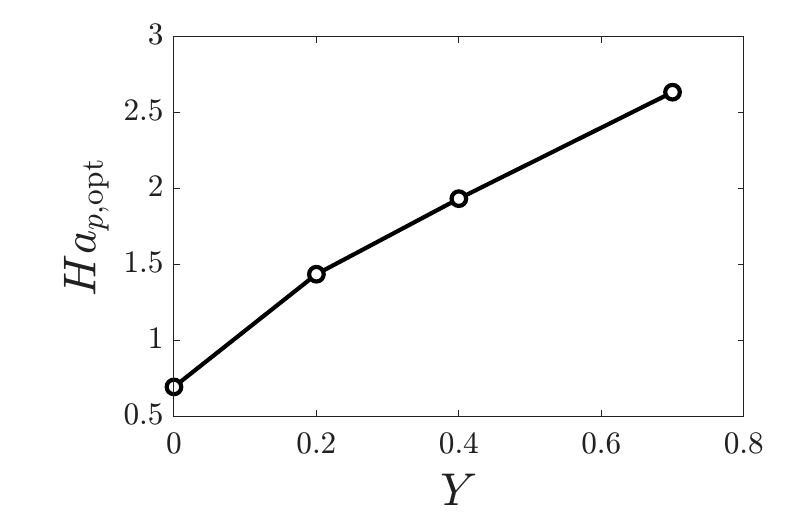}
\caption{}\label{fig:actual_peak_location_new}
\end{subfigure}
\par\smallskip
\begin{subfigure}{0.49\linewidth}
\centering
\includegraphics[width=\linewidth]{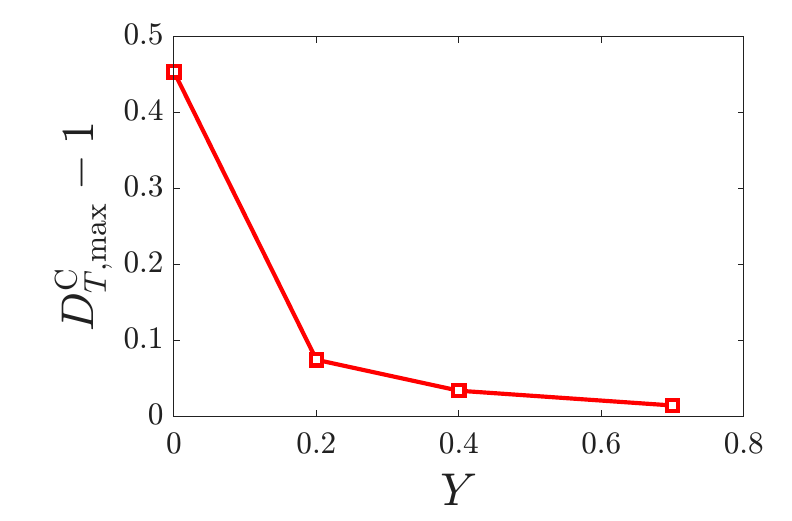}
\caption{}\label{fig:actual_peak_amplitude_new}
\end{subfigure}\hfill
\begin{subfigure}{0.49\linewidth}
\centering
\includegraphics[width=\linewidth]{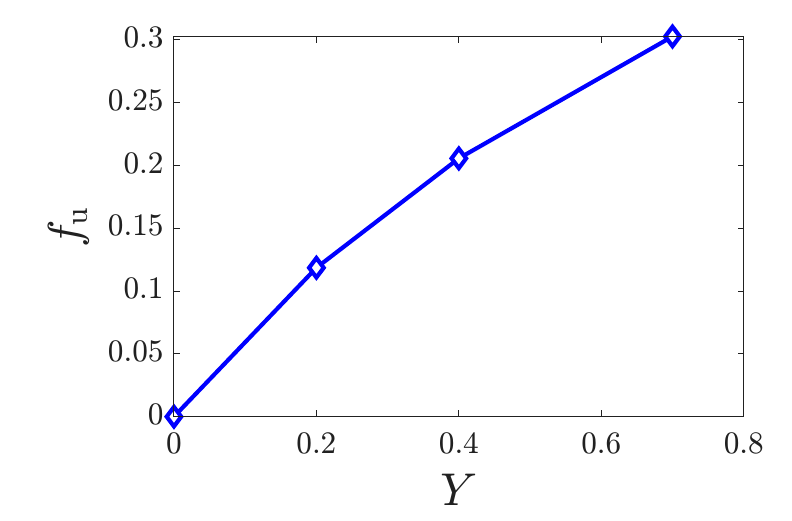}
\caption{}\label{fig:actual_peak_yield_new}
\end{subfigure}
\caption{Casson dispersion under fixed forcing: (a) $D_T^{\mathrm C}(Ha_p)$ and computed optima; (b) $Ha_{p,\mathrm{opt}}$; (c) $D_{T,\mathrm{max}}^{\mathrm C}-1$; and (d) unyielded fraction at the optimum. Parameters are given in Table~\ref{tab:true_casson_peaks_new}. Symbols in (b)--(d) mark the computed yield-stress cases.}
\label{fig:true_casson_optima_new}
\end{figure*}

Table~\ref{tab:true_casson_peaks_new} and Fig.~\ref{fig:true_casson_optima_new} show the systematic displacement and attenuation of the magnetic dispersion peak as yielding increases. Over $0\leq Ha_p\leq4$, raising $Y$ from $0$ to $0.7$ shifts the optimum from $Ha_p=0.69554$ to $2.63351$ and reduces the peak enhancement $D_{T,\mathrm{max}}^{\mathrm C}-1$ from $0.45358$ to $0.01469$. The unyielded fraction grows to about $0.303$, while the mean velocity at the optimized states decreases from approximately $0.566$ to $0.161$. The magnetic optimum consequently moves as the yielding threshold changes the stress distribution, the velocity in the sheared layers and the total flow. Unlike the Newtonian electromagnetic response~\cite{Hossain2025}, its location and amplitude now depend on an evolving yield surface. Equation~\eqref{eq:casson_cumulative_variation_new} relates the reduction in dispersion to the cumulative velocity departure across both the yielded and unyielded portions of the slit.

\subsection{Dispersion at fixed mean throughput}\label{sec:equal_throughput_new}

To distinguish the influence of rheology from that of a changing flow rate, we prescribe the dimensionless mean velocity $\langle\widehat u\rangle=0.08$ for each fluid and each $Ha_p$. For the fixed slit geometry, this is equivalent to prescribing the dimensional volumetric flow rate $Q=2hW\langle u^*\rangle=2hWU_0\langle\widehat u\rangle$, where $W$ is the channel width. The pressure parameter $\Pi$ is then calculated from the momentum equation, keeping $\Omega=0.12$, $S_p=1$, $\kappa=10$ and $Pe_0=10$ fixed. For comparison, a linear-viscosity fluid with $\beta=3$ is evaluated at the same mean throughput and viscosity scale $\mu=\mu_p$.

\begin{figure*}
\centering
\begin{subfigure}{0.49\linewidth}
\centering
\includegraphics[width=\linewidth]{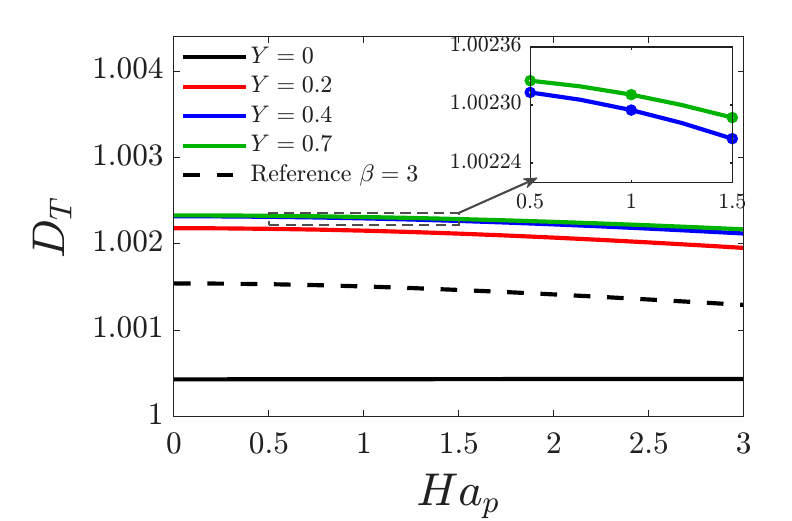}
\caption{}\label{fig:matched_dispersion_new}
\end{subfigure}\hfill
\begin{subfigure}{0.49\linewidth}
\centering
\includegraphics[width=\linewidth]{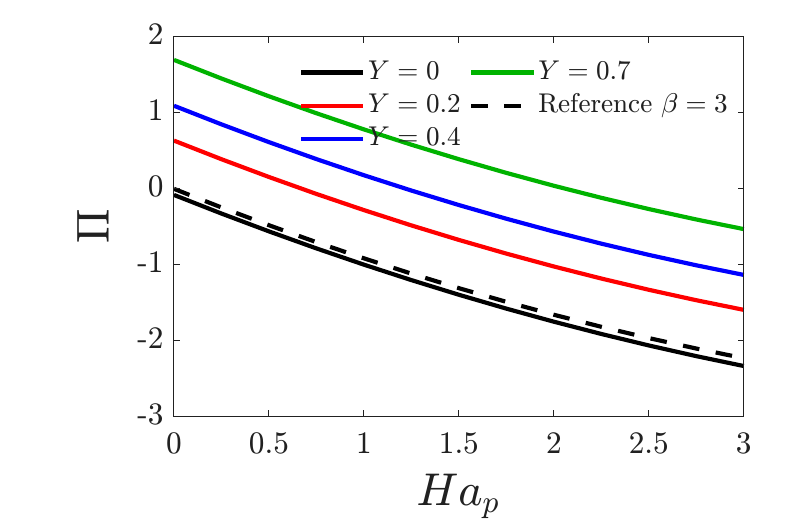}
\caption{}\label{fig:matched_pressure_new}
\end{subfigure}
\par\smallskip
\begin{subfigure}{0.49\linewidth}
\centering
\includegraphics[width=\linewidth]{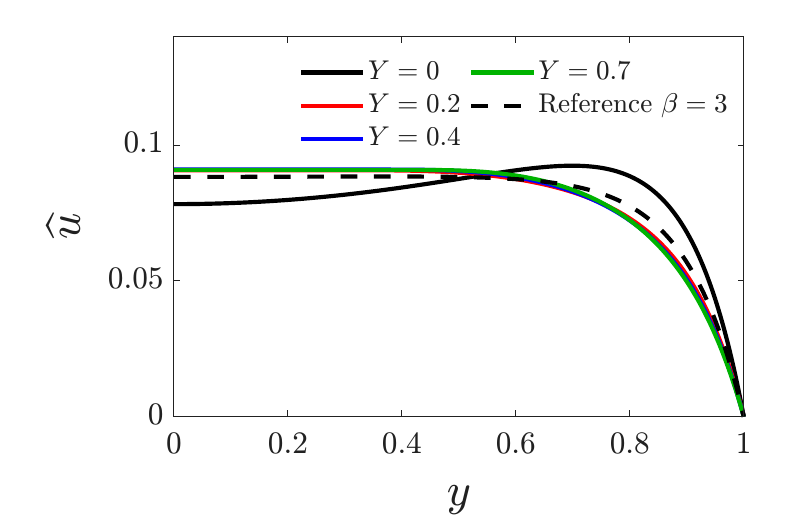}
\caption{}\label{fig:matched_velocity_new}
\end{subfigure}\hfill
\begin{subfigure}{0.49\linewidth}
\centering
\includegraphics[width=\linewidth]{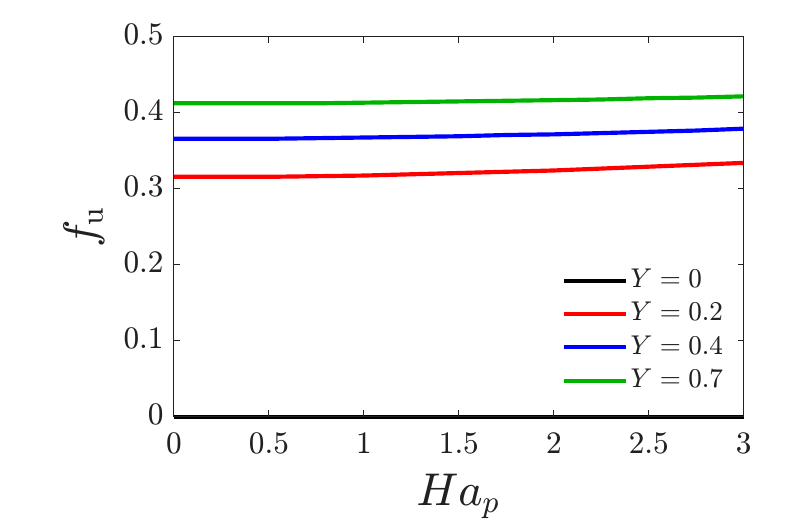}
\caption{}\label{fig:matched_yield_new}
\end{subfigure}
\caption{Fixed-mean-throughput comparison at $\langle\widehat u\rangle=0.08$: (a) normalized dispersion with a magnified inset; (b) required pressure forcing; (c) velocity at $Ha_p=1$; and (d) unyielded fraction. The black dashed curve denotes the effective-viscosity reference ($\beta=3$).}
\label{fig:matched_throughput_new}
\end{figure*}

\begin{table}
\caption{Representative fixed-mean-throughput solutions at $Ha_p=1$, with $\langle\widehat u\rangle=0.08$, $Pe_0=10$ and all other dimensionless parameters as in Fig.~\ref{fig:matched_throughput_new}. The required pressure forcing is allowed to change sign.}
\label{tab:matched_throughput_values_new}
\centering
\begin{tabular}{lrr}
\toprule
Material & Required $\Pi$ & $D_{\mathrm{eff}}/D$ \\
\midrule
Casson $Y=0$ & $-1.00081$ & $1.000433$ \\
Casson $Y=0.2$ & $-0.28334$ & $1.002153$ \\
Casson $Y=0.4$ & $\phantom{-}0.17518$ & $1.002295$ \\
Casson $Y=0.7$ & $\phantom{-}0.77805$ & $1.002311$ \\
Effective viscosity $\beta=3$ & $-0.91769$ & $1.001507$ \\
\bottomrule
\end{tabular}
\end{table}

At a prescribed mean throughput, the dependence on yielding differs sharply from the fixed-forcing response (Fig.~\ref{fig:matched_throughput_new}). The pressure parameter now adjusts with $Y$ to maintain $\langle\widehat u\rangle=0.08$ while the electrokinetic and magnetic forcing parameters remain fixed. Consequently, the mean advection is identical but the velocity deviations entering the homogenized auxiliary problem are different. At $Ha_p=1$, $D_T^{\mathrm C}$ increases from $1.000433$ at $Y=0$ to $1.002311$ at $Y=0.7$ (Table~\ref{tab:matched_throughput_values_new}). The shear-induced enhancement $D_T^{\mathrm C}-1$ therefore grows by a factor of approximately $5.34$, while the total normalized diffusivity increases by approximately $0.188\%$. The required pressure gradient changes sign for some cases and redistributes the velocity differences between the central plug and near-wall layers. This is the mechanistic origin of the reversed trend. Classical equal-flux Casson dispersion was examined by Dash \emph{et al.}~\cite{DashJayaramanMehta2000}; here the compensating pressure gradient acts simultaneously with maintained electroosmotic and Lorentz forcing. The choice between pressure regulation and flow-rate control can therefore reverse the dispersion response to an increase in yield stress. For microfluidic sample transport, this distinction is relevant when interpreting how a change in suspension rheology affects longitudinal band broadening.

\subsection{Numerical benchmarks and convergence}\label{sec:casson_numeric_validation_new}

\begin{figure}[t]
\centering
\includegraphics[width=0.5\linewidth]{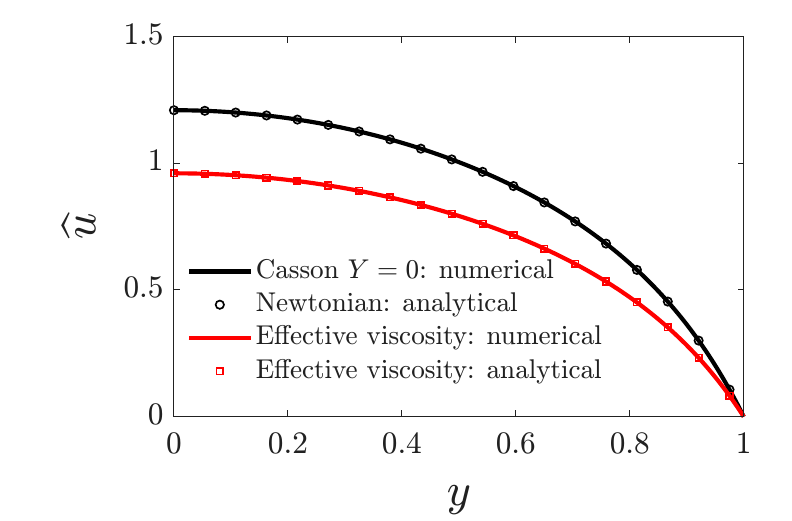}
\caption{Velocity-profile validation. The genuine Casson solution at $Y=0$ recovers the Newtonian analytical profile, while the effective-viscosity reference solution at $\beta=3$ agrees with its corresponding analytical profile. Lines denote numerical results and symbols denote the analytical solutions. Parameters: $\Pi=0.7$, $\Omega=0.4$, $S_p=2$, $Ha_p=0.8$, and $\kappa=7$.}
\label{fig:newtonian_reference_velocity_checks_new}
\end{figure}

The validation is organized to test the velocity calculation and the homogenized transport coefficient independently. The Casson stress--velocity solver reproduces the exact pressure-driven profiles~\eqref{eq:casson_exact} for $\alpha=0,0.2,0.5,0.7,0.85$ with a maximum velocity error of $1.02\times10^{-8}$. In the Newtonian and linear-viscosity limits, numerical velocities agree with the corresponding analytical solutions within $5.58\times10^{-10}$ and $4.32\times10^{-10}$ (Fig.~\ref{fig:newtonian_reference_velocity_checks_new}). The plane-Poiseuille auxiliary problem recovers the Taylor--Aris coefficient $2/945$ using $U_P=Gh^2/\mu_p$. The numerically computed first corrector also satisfies the homogeneous Neumann boundary conditions and the homogenization energy identity
\begin{equation}\label{eq:auxiliary_energy_num_new}
\left\langle\left(\frac{\dd\widehat A_1}{\dd y}\right)^2\right\rangle
=-\langle\widehat u\widehat A_1\rangle,\qquad
\left.\frac{\dd\widehat A_1}{\dd y}\right|_{y=\pm1}=0.
\end{equation}

\begin{table*}
\caption{Transverse-grid convergence of the nonlinear Casson velocity and Taylor--Aris auxiliary boundary-value problem at $Y=0.4$, $Ha_p=1.91$, $\Pi=1$, $\Omega=0.12$, $S_p=1$, $\kappa=10$, $Pe_0=10$. The reported difference is relative to the $N=3201$ reference solution; its own difference is exactly zero by definition, not a continuum-error estimate.}
\label{tab:true_casson_grid_new}
\centering
\begin{tabular}{rrrr}
\toprule
Grid points $N$ & $D_T^{\mathrm C}$ & $|D_T^{(N)}-D_T^{(3201)}|$ & Yield-front position \\
\midrule
$201$  & $1.03412764658$ & $3.7202\times10^{-6}$ & $0.20547822$ \\
$401$  & $1.03412484555$ & $9.1914\times10^{-7}$ & $0.20547824$ \\
$801$  & $1.03412414525$ & $2.1884\times10^{-7}$ & $0.20547824$ \\
$1601$ & $1.03412397017$ & $4.3769\times10^{-8}$ & $0.20547824$ \\
$3201$ & $1.03412392641$ & $0$ (reference) & $0.20547825$ \\
\bottomrule
\end{tabular}
\end{table*}

Table~\ref{tab:true_casson_grid_new} demonstrates transverse-grid convergence at $Y=0.4$ and $Ha_p=1.91$. Increasing the grid resolution from $N=1601$ to $3201$ changes $D_T^{\mathrm C}$ by $4.38\times10^{-8}$. The wall-velocity and auxiliary-function Neumann boundary residuals remain close to machine precision at this parameter set, indicating that the reported long-time coefficient is well resolved.

\subsection{Concentration fields in the linear-viscosity model}

\begin{figure*}
	\centering
	\begin{subfigure}{0.45\linewidth}
		\centering
		\includegraphics[width=\linewidth]{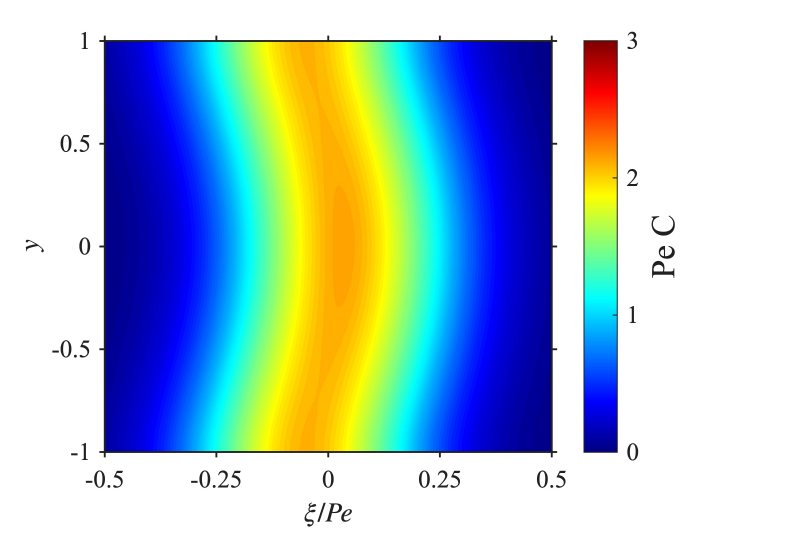}
		\caption{}
		\label{fig:4aa}
	\end{subfigure}
		\begin{subfigure}{0.45\linewidth}
		\centering
		\includegraphics[width=\linewidth]{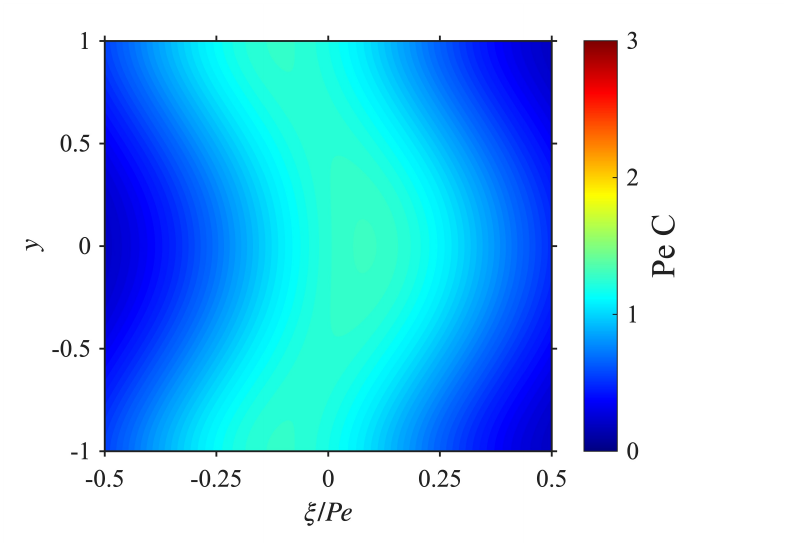}
		\caption{}
		\label{fig:4ab}
	\end{subfigure}
		\begin{subfigure}{0.45\linewidth}
		\centering
		\includegraphics[width=\linewidth]{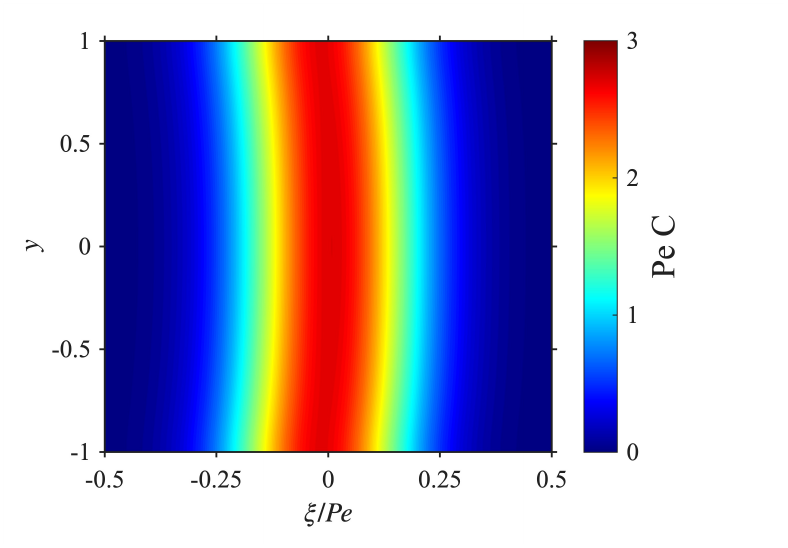}
		\caption{}
		\label{fig:4ac}
	\end{subfigure}
	\begin{subfigure}{0.45\linewidth}
		\centering
		\includegraphics[width=\linewidth]{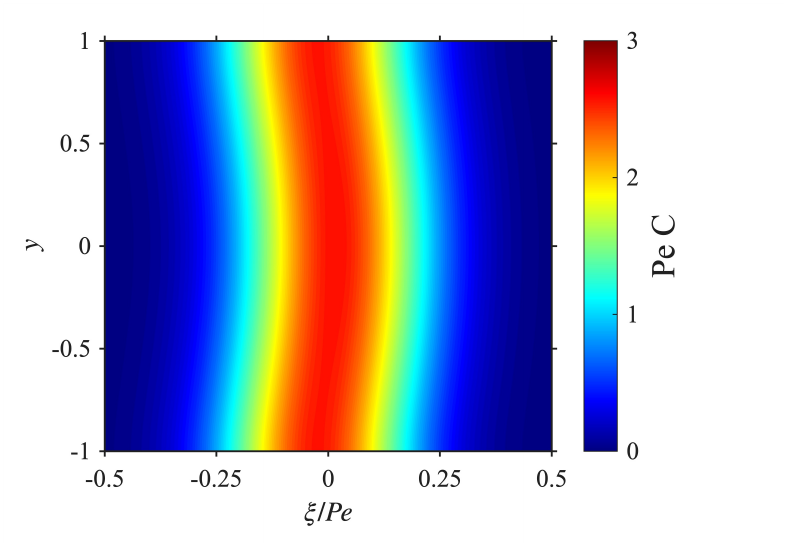}
		\caption{}
		\label{fig:4ad}
	\end{subfigure}
	\begin{subfigure}{0.45\linewidth}
		\centering
		\includegraphics[width=\linewidth]{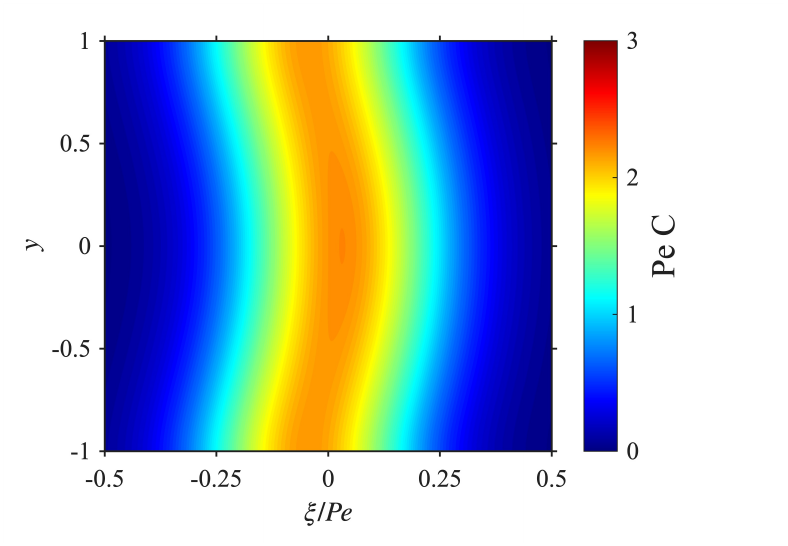}
		\caption{}
		\label{fig:4ae}
	\end{subfigure}
		\begin{subfigure}{0.45\linewidth}
		\centering
		\includegraphics[width=\linewidth]{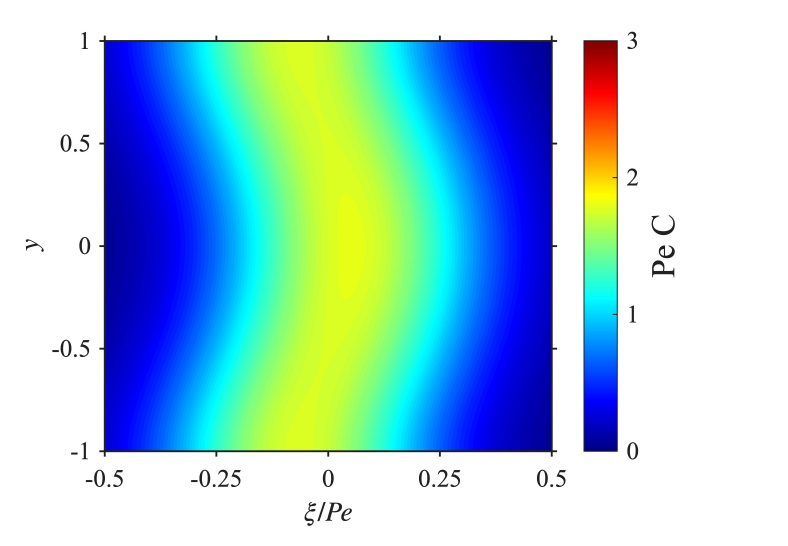}
		\caption{}
		\label{fig:4af}
	\end{subfigure}
	\caption{Concentration contours $Pe\,C$ reconstructed through third order in the Taylor--Aris expansion for the analytical linear-viscosity velocity: (a) $\beta\to\infty$, $Ha=0$; (b) $\beta\to\infty$, $Ha=1$; (c) $\beta\to\infty$, $Ha=6$; (d) $\beta=1$, $Ha=0$; (e) $\beta=10$, $Ha=0$; and (f) $\beta=1$, $Ha=1$. Unless varied, $\beta_1=1$, $S=5$, $\kappa=10$, $Pe=10$, and $\tau=1$. The abscissa is $\xi/Pe$, and all colourbars display $Pe\,C$ on the common interval $[0,3]$. These are finite-order asymptotic reconstructions, not direct two-dimensional finite-time PDE solutions.}
	\label{fig:4aaa}
\end{figure*}

The concentration fields in Fig.~\ref{fig:4aaa} use the analytical linear-viscosity velocity~\eqref{eq:16}. Magnetic forcing changes the carrier-flow shear and hence the width and peak amplitude of the tracer cloud; the neutral tracer experiences no direct magnetic force. Symmetric wall conditions and release make the fields even in $y$, independently of $\beta$. With $q=\xi/Pe$, the contours display $Pe\,C(Pe\,q,y,\tau)$, so that the integrated concentration is preserved under the change of axial coordinate. They are third-order reconstructions of a unit-mass impulse in the long-time Taylor--Aris expansion, rather than direct finite-time solutions of the two-dimensional transport equation.

\subsection{Transverse concentration variation}

\begin{figure*}
	\centering
	\begin{subfigure}{0.47\linewidth}
		\centering
		\includegraphics[width=\linewidth]{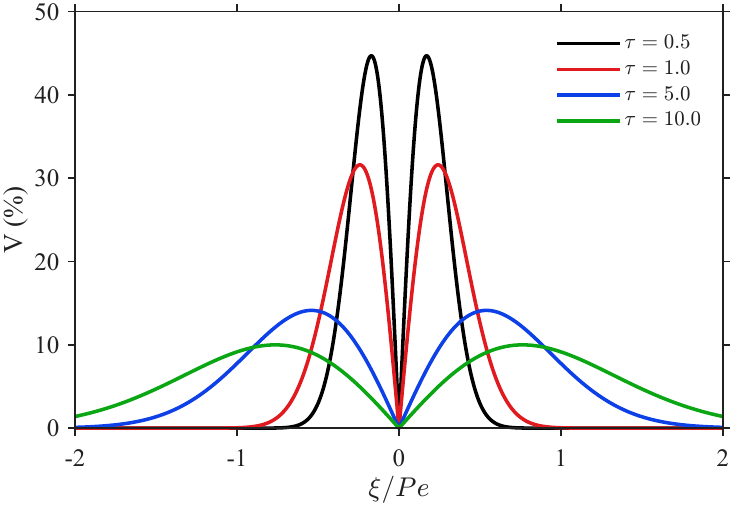}
		\caption{}
		\label{fig:12a}
	\end{subfigure}
	\begin{subfigure}{0.47\linewidth}
		\centering
		\includegraphics[width=\linewidth]{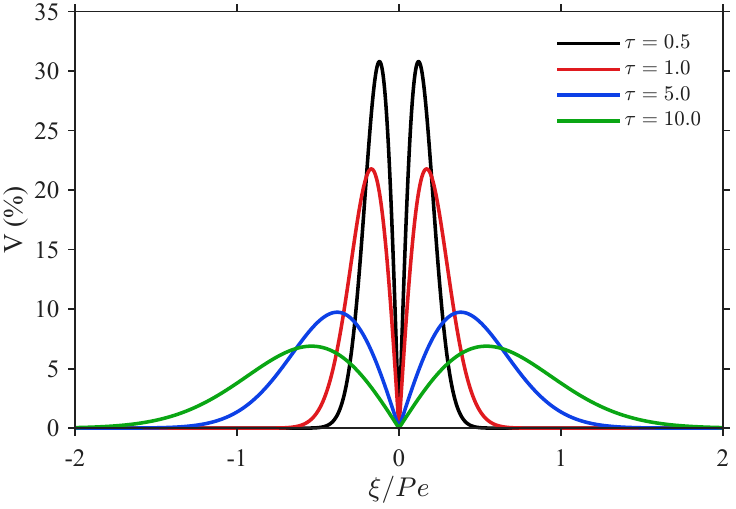}
		\caption{}
		\label{fig:12b}
	\end{subfigure}
	\begin{subfigure}{0.47\linewidth}
		\centering
		\includegraphics[width=\linewidth]{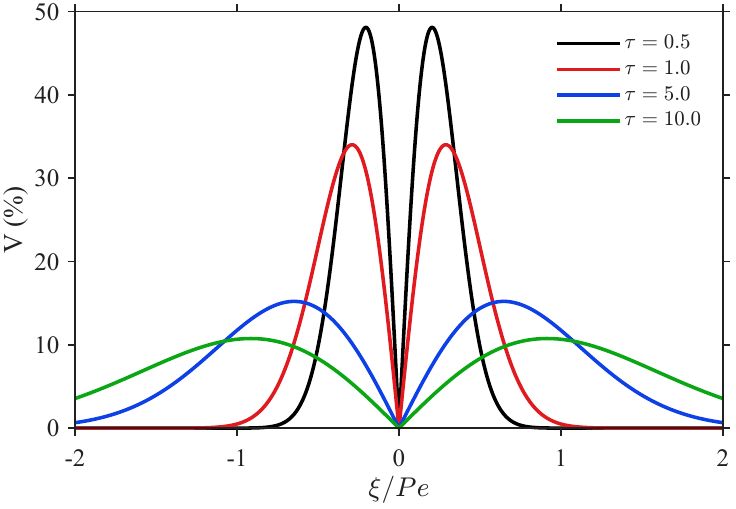}
		\caption{}
		\label{fig:12c}
	\end{subfigure}
	\begin{subfigure}{0.47\linewidth}
		\centering
		\includegraphics[width=\linewidth]{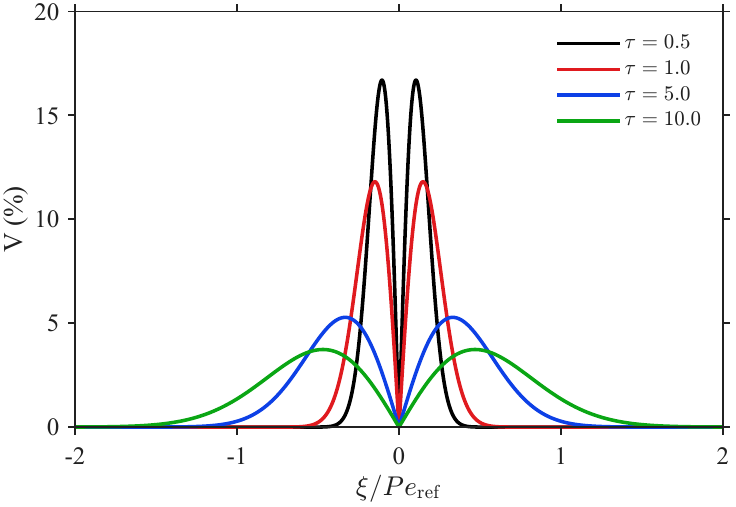}
		\caption{}
		\label{fig:12d}
	\end{subfigure}
	\begin{subfigure}{0.47\linewidth}
		\centering
		\includegraphics[width=\linewidth]{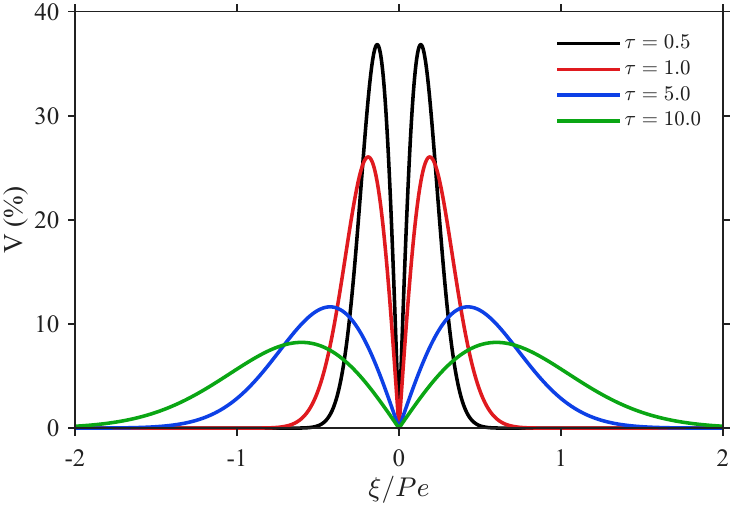}
		\caption{}
		\label{fig:12e}
	\end{subfigure}
	\begin{subfigure}{0.47\linewidth}
		\centering
		\includegraphics[width=\linewidth]{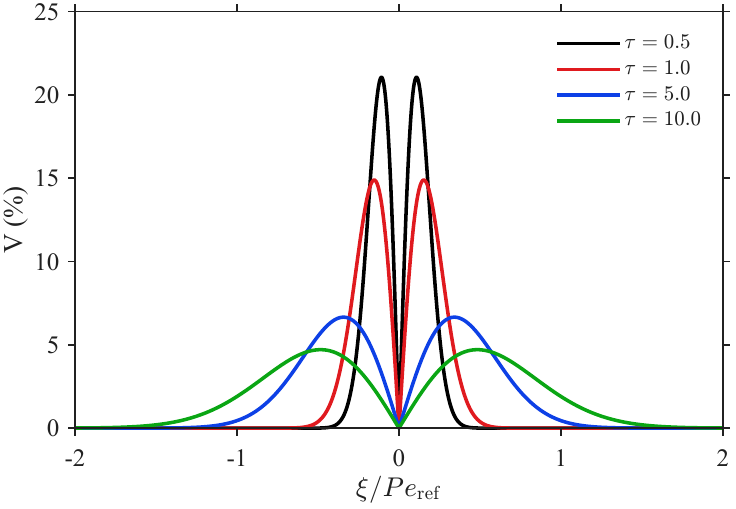}
		\caption{}
		\label{fig:12f}
	\end{subfigure}
	
	\caption{First-order approximation to the transverse concentration-range measure $V$ (\%) versus the scaled axial coordinate at $\tau=0.5$, $1$, $5$ and $10$: (a) $\beta=3$, $Ha=0.5$; (b) $\beta=3$, $Ha=0$; (c) $\beta\to\infty$, $Ha=0.5$; (d) pressure-driven magnetic-braking reference with $\beta=3$, $Ha=0.5$; (e) $\beta\to\infty$, $Ha=0$; and (f) pressure-driven magnetic-braking reference with $\beta\to\infty$, $Ha=0.5$. Panels (a)--(c) and (e) use $\beta_1=1$, $S=5$, $\kappa=10$, and $Pe=10$, with abscissa $\xi/Pe$. Panels (d) and (f) neglect externally applied axial and spanwise electric forcing, use $\Pi_{\mathrm{ref}}=1$ and $Pe_{\mathrm{ref}}=10$, and show $\xi/Pe_{\mathrm{ref}}$ [Eqs.~\eqref{eq:pressure_only_scaling}--\eqref{eq:pressure_only_exact}]. The ordinate is normalized by the centre of the leading Gaussian pulse; the curves are not the full finite-time transverse range.}
	\label{fig:12}
\end{figure*}

We characterize cross-sectional concentration nonuniformity by the percentage range measure~\cite{wu14},
\begin{equation}\label{eq:V_definition}
V(\xi,\tau)=100\,\frac{\displaystyle\max_{-1\leq y\leq1}C(\xi,y,\tau)-\displaystyle\min_{-1\leq y\leq1}C(\xi,y,\tau)}{C(0,0,\tau)}.
\end{equation}
Here $V$ is a transverse range relative to the concentration at the channel centre and moving-frame origin, rather than a statistical variance. In Fig.~\ref{fig:12}, we evaluate this diagnostic using only the first transverse correction in the Taylor--Aris expansion, so that the field used in the figure is
\begin{equation}\label{eq:first_order_reconstructed_C}
 C_{\mathrm{tr}}(\xi,y,\tau)=C_0(\xi,\tau)
 +Pe\,A_1(y)\frac{\partial C_0}{\partial\xi}.
\end{equation}
For this approximation, the denominator in Eq.~\eqref{eq:V_definition} is replaced consistently by $C_0(0,\tau)$; the zero-mean auxiliary function ensures that this is also the cross-sectional mean of $C_{\mathrm{tr}}$ at the plume centre. Writing $\Delta A_1=\max_{-1\leq y\leq1} A_1(y)-\min_{-1\leq y\leq1} A_1(y)$ and using the Gaussian~\eqref{eq:80}, we obtain the explicit first-order approximation
\begin{equation}\label{eq:first_order_V_formula}
 V(\xi,\tau)\simeq100\,|Pe|\,\Delta A_1
 \frac{|\xi|}{2D_T\tau}
 \exp\!\left(-\frac{\xi^2}{4D_T\tau}\right).
\end{equation}
For the pressure-driven reference panels, the same expression is evaluated using $Pe_{\mathrm{ref}}$, the associated auxiliary function and the pressure-only velocity~\eqref{eq:pressure_only_exact}. The first-order approximation vanishes at $\xi=0$, since the longitudinal derivative of a symmetric Gaussian vanishes there, and has paired maxima at $|\xi|=\sqrt{2D_T\tau}$. This accounts for the characteristic curves in Fig.~\ref{fig:12}. It does not imply that the transverse range of the full concentration field is zero at the plume centre: the next correction involves $A_2\,\partial^2 C_0/\partial\xi^2$ and generally contributes there. Moreover, the asymptotic reconstruction does not uniformly describe the initial transverse-relaxation interval.

\FloatBarrier
\subsection{Finite-time tracer transport}\label{sec:direct_2d_casson_new}

\begin{figure*}[t]
\centering
\begin{subfigure}[t]{0.49\linewidth}
\centering
\includegraphics[width=\linewidth]{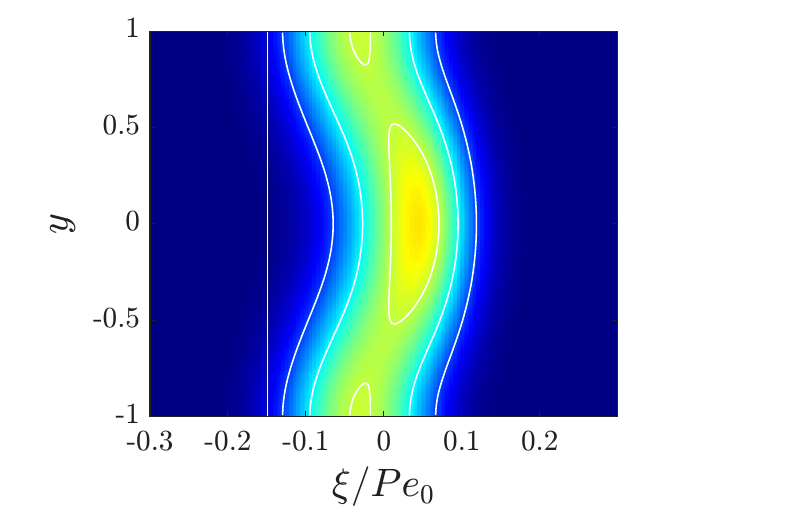}
\caption{}\label{fig:casson_conc_y0_new}
\end{subfigure}\hfill
\begin{subfigure}[t]{0.49\linewidth}
\centering
\includegraphics[width=\linewidth]{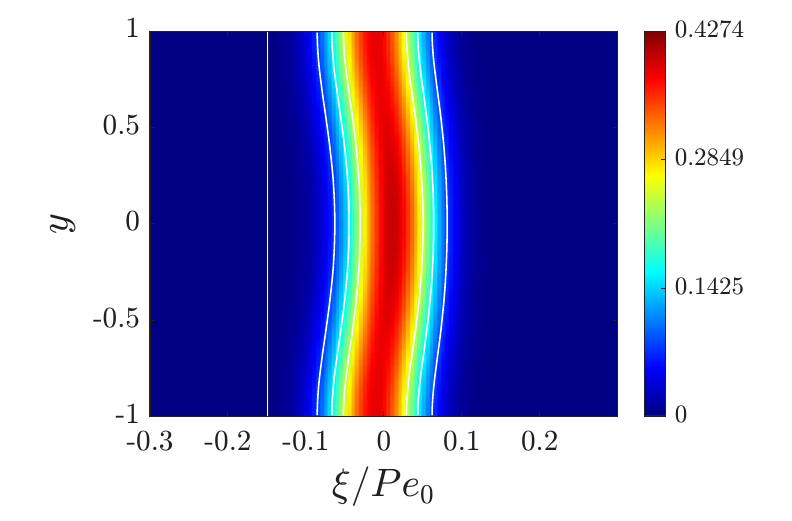}
\caption{}\label{fig:casson_conc_y02_new}
\end{subfigure}
\par\smallskip
\begin{subfigure}[t]{0.49\linewidth}
\centering
\includegraphics[width=\linewidth]{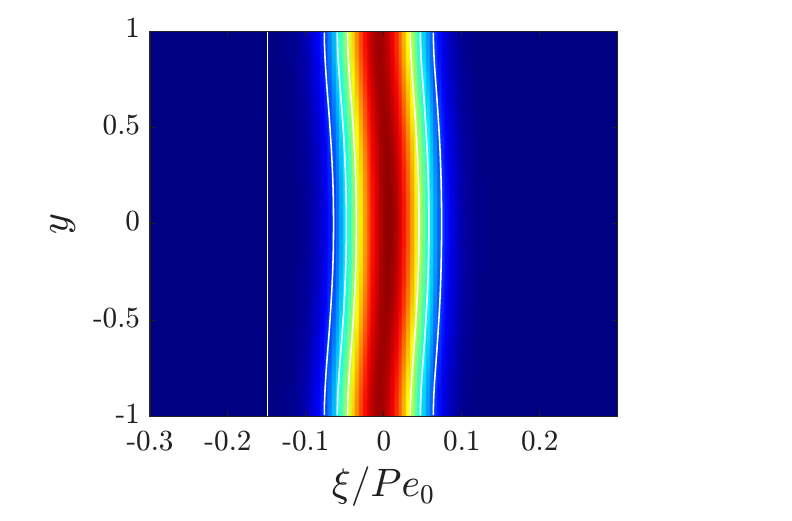}
\caption{}\label{fig:casson_conc_y04_new}
\end{subfigure}\hfill
\begin{subfigure}[t]{0.49\linewidth}
\centering
\includegraphics[width=\linewidth]{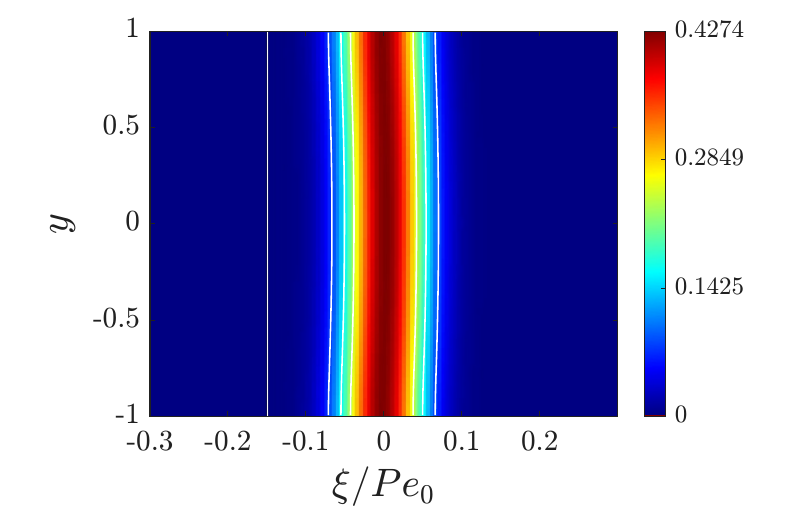}
\caption{}\label{fig:casson_conc_y07_new}
\end{subfigure}
\caption{Direct dimensionless concentration $C$ at $\tau=0.4$ for (a) $Y=0$, (b) $Y=0.2$, (c) $Y=0.4$, and (d) $Y=0.7$. The four panels share one concentration range; white contours denote $C=0.08,0.16,0.24$. Here $Pe_0=25$, $\sigma_0=0.25$, $Ha_p=0.8$, $\Pi=1$, $\Omega=0.12$, $S_p=1$ and $\kappa=10$.}
\label{fig:direct_casson_concentrations_new}
\end{figure*}

The long-time coefficient obtained by homogenization is examined against the evolving tracer plume by solving the full two-dimensional advection--diffusion equation~\eqref{eq:17} with the computed Casson velocity. This direct calculation tests the time-dependent transport separately from the asymptotic auxiliary problem. The numerical initial condition is a unit-mass Gaussian pulse, uniform across the channel section and specified in the moving frame by
\begin{equation}\label{eq:direct_gaussian_initial}
C(\xi,y,0)=\frac{1}{\sqrt{2\pi}\,\sigma_0}
\exp\!\left(-\frac{\xi^2}{2\sigma_0^2}\right),
\qquad -1\leq y\leq1.
\end{equation}
Here $\sigma_0$ is the dimensionless initial axial standard deviation measured in units of $h$; the pulse is distinct from the idealized Dirac impulse~\eqref{eq:18}. Two explicitly distinguished parameter sets are used below: $Pe_0=25$ and $\sigma_0=0.25$ for the concentration comparison in Fig.~\ref{fig:direct_casson_concentrations_new}, and $Pe_0=10$ and $\sigma_0=0.4$ for the signed-departure and numerical-verification calculations. The concentration is advanced in the moving coordinate $\xi=x^*/h-Pe_0\langle\widehat u\rangle\tau$ with impermeable walls, using the discretization described in Sec.~\ref{sec:numerical_methods}.

The direct solutions in Fig.~\ref{fig:direct_casson_concentrations_new} reveal an appreciable rheological dependence when the advective scale is increased to $Pe_0=25$ and the initially narrower Gaussian tracer pulse is observed at $\tau=0.4$. For the same imposed pressure and electromagnetic forcing, the plume is substantially broader when $Y=0$ than when $Y=0.7$. The computed axial variance decreases from approximately $2.586$ to $0.874$, while the peak concentration rises from $0.276$ to $0.427$ across these two cases. The white concentration isocontours provide a common geometric reference for comparing the widths and transverse distortion of the four plumes. The narrower band at larger $Y$ reflects the reduced velocity variation across the section; plug width alone does not determine that variation. The first homogenized auxiliary problem measures the cumulative difference between the local velocity and the mean. These fields are obtained from the direct time-dependent transport solver at the stated dimensionless forcing, rather than from an assumption of experimentally attainable magnetic fields.

To resolve the finite-time deviations from the long-time Taylor--Aris description, we compare the direct concentration with the leading homogenized Gaussian concentration for the \emph{same} initial tracer pulse. Its variance includes the initially prescribed width:
\begin{equation}\label{eq:gaussian_c0_finitewidth_new}
C_0(\xi,\tau;\sigma_0)=
\frac{\exp\!\left[-\xi^2/(2w(\tau))\right]}{\sqrt{2\pi w(\tau)}},
\qquad w(\tau)=\sigma_0^2+2D_T^{\mathrm C}\tau.
\end{equation}
In Eq.~\eqref{eq:gaussian_c0_finitewidth_new}, $C_0$ is the cross-sectionally uniform leading-order homogenized concentration initialized with the same Gaussian tracer pulse, whereas $C(\xi,y,\tau)$ is the full two-dimensional numerical field. The actual cross-sectional mean is $\langle C\rangle(\xi,\tau)=\frac12\int_{-1}^1 C(\xi,y,\tau)\,\dd y$ [cf. Eq.~\eqref{eq:37}]. The asymptotic normalization $\langle C_1\rangle=\langle C_2\rangle=\langle C_3\rangle=0$ gives $\langle C\rangle=C_0+O(\varepsilon^4)$ in the formal long-time expansion; it does not imply that the full finite-time solution has exactly the Gaussian cross-sectional mean. Accordingly, $C-C_0$ measures departure from the homogenized prediction and is generally different from $C-\langle C\rangle$. Indeed, $\langle C-\langle C\rangle\rangle=0$ by definition, whereas $\langle C-C_0\rangle=\langle C\rangle-C_0$ generally differs from zero at finite time.

\begin{figure}[!htbp]
\centering
\begin{subfigure}[t]{0.49\linewidth}
\centering
\includegraphics[width=\linewidth]{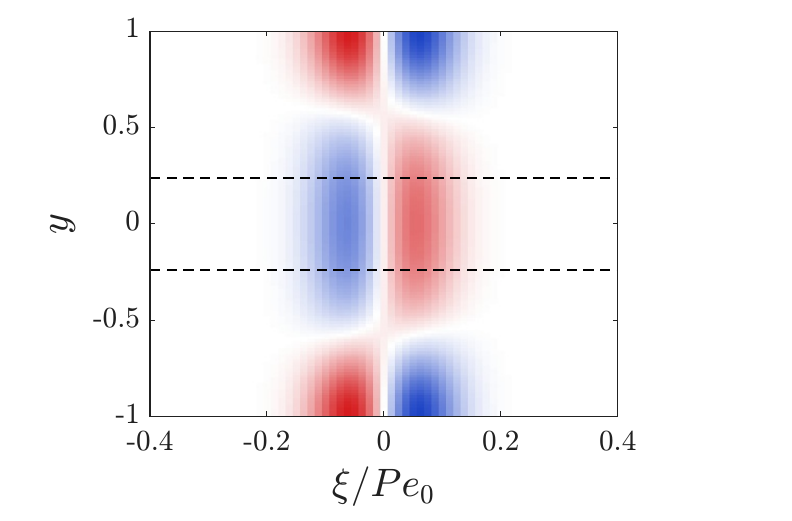}
\caption{}\label{fig:dep_y04_t01_new}
\end{subfigure}\hfill
\begin{subfigure}[t]{0.49\linewidth}
\centering
\includegraphics[width=\linewidth]{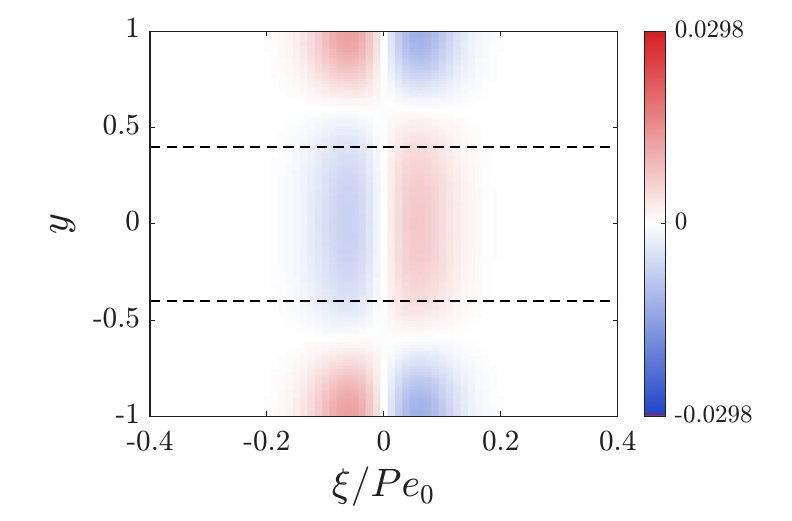}
\caption{}\label{fig:dep_y07_t01_new}
\end{subfigure}
\par\smallskip
\begin{subfigure}[t]{0.49\linewidth}
\centering
\includegraphics[width=\linewidth]{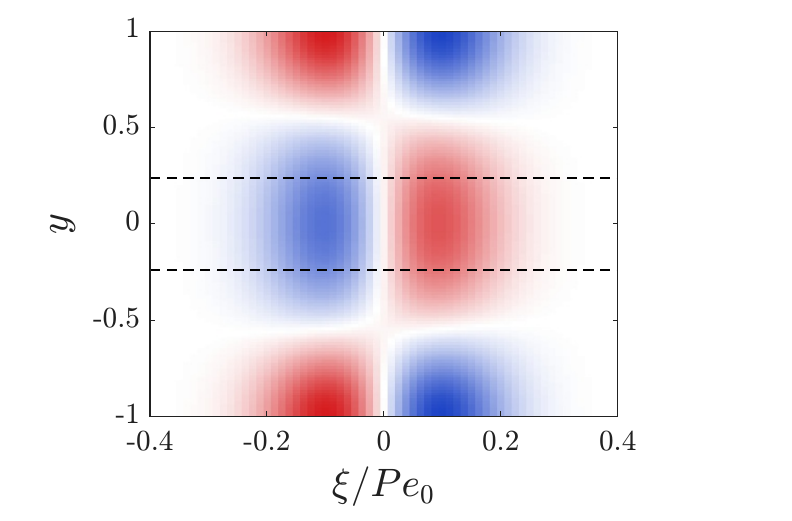}
\caption{}\label{fig:dep_y04_t04_new}
\end{subfigure}\hfill
\begin{subfigure}[t]{0.49\linewidth}
\centering
\includegraphics[width=\linewidth]{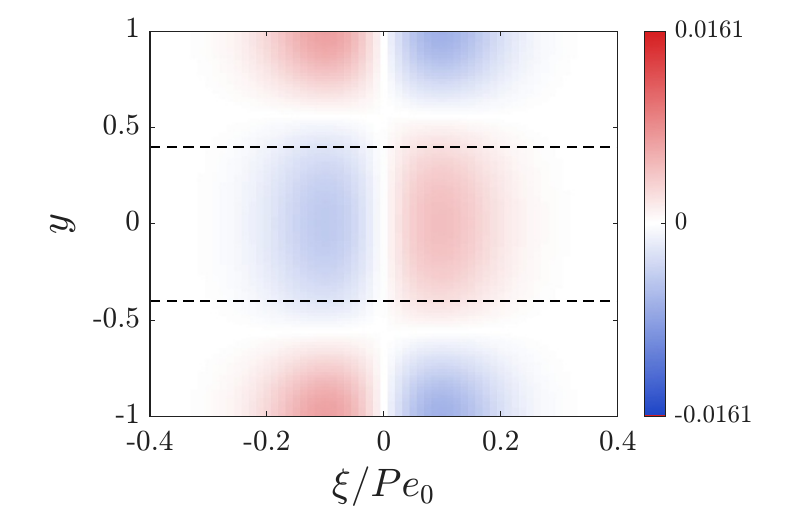}
\caption{}\label{fig:dep_y07_t04_new}
\end{subfigure}
\par\smallskip
\begin{subfigure}[t]{0.49\linewidth}
\centering
\includegraphics[width=\linewidth]{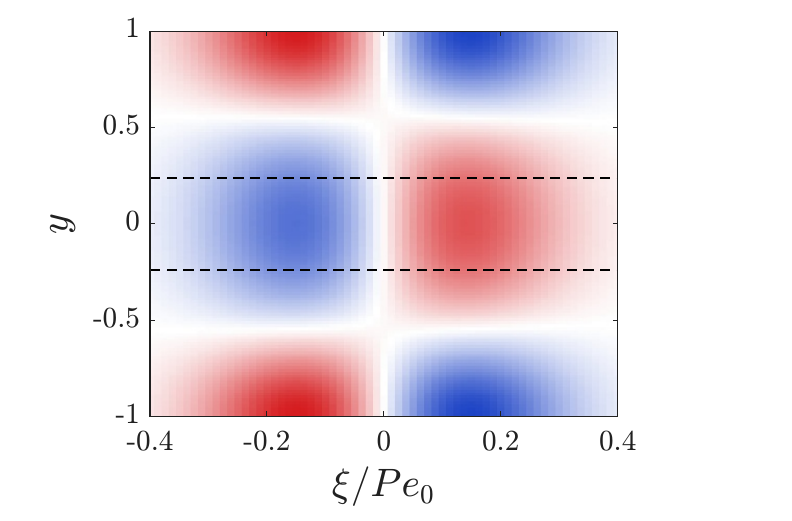}
\caption{}\label{fig:dep_y04_t10_new}
\end{subfigure}\hfill
\begin{subfigure}[t]{0.49\linewidth}
\centering
\includegraphics[width=\linewidth]{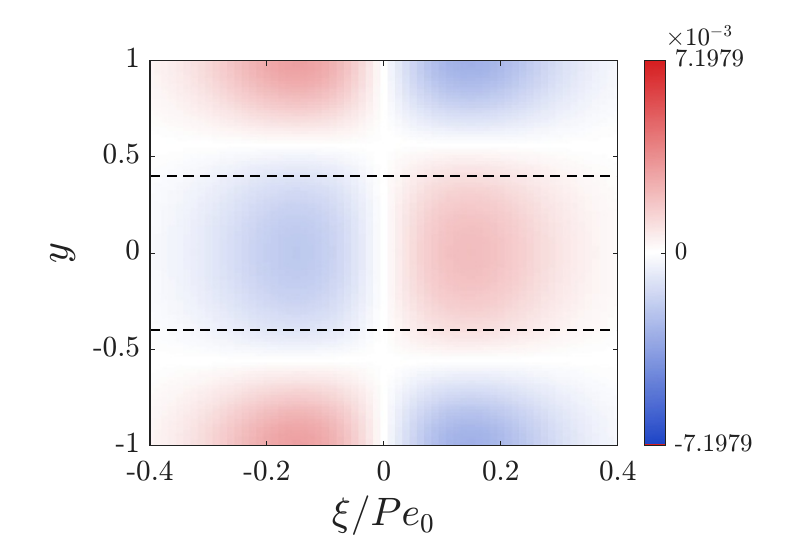}
\caption{}\label{fig:dep_y07_t10_new}
\end{subfigure}
\caption{Signed finite-time concentration departure $C-C_0$ for $Y=0.4$ (left) and $Y=0.7$ (right), at (a,b) $\tau=0.1$, (c,d) $\tau=0.4$ and (e,f) $\tau=1$. The colour range is symmetric and common to both panels at each time, but changes between times. Dashed lines mark the yield fronts; $Pe_0=10$ and $\sigma_0=0.4$.}
\label{fig:casson_departure_new}
\end{figure}

Figure~\ref{fig:casson_departure_new} resolves the signed departure from the finite-width homogenized Gaussian. In the plug, local advection is uniform, but the plug velocity generally differs from the cross-sectional mean; this difference produces finite-time concentration distortion despite a vanishing local shear rate. The adjoining yielded layers contribute through their continuous velocity gradients. The opposing red and blue lobes consequently reflect axial displacement relative to the reference Gaussian rather than wall uptake or an additional source of solute. They remain symmetric under reflection about $y=0$, consistent with the symmetric boundaries and forcing.

The maximum absolute departure for $Y=0.4$ decreases from approximately $2.98\times10^{-2}$ at $\tau=0.1$ to $1.61\times10^{-2}$ at $\tau=0.4$ and $7.20\times10^{-3}$ at $\tau=1$. For $Y=0.7$, the corresponding values are $1.25\times10^{-2}$, $6.67\times10^{-3}$ and $3.00\times10^{-3}$. Thus the maximum departure decays over the tested observation times and is lower in the more strongly yielded case under the same forcing. Because the colourbar limits are selected separately for each observation time, the quantitative values, rather than the visual saturation of different rows, must be used for comparisons across time. These finite-time results are consistent with decreasing velocity heterogeneity, but are not a proof of uniform validity of the long-time expansion at $\tau\ll1$.

An independent calculation of the transient axial moments follows from the hierarchy $m_n(y,\tau)=\int\xi^n C(\xi,y,\tau)\,\dd\xi$:
\begin{equation}\label{eq:direct_moments_new}
\frac{\partial m_n}{\partial\tau}=\frac{\partial^2m_n}{\partial y^2}
+nPe_0\bigl(\widehat u-\langle\widehat u\rangle\bigr)m_{n-1}
+n(n-1)m_{n-2},\qquad n=0,1,2,\ldots,
\end{equation}
For negative indices we set $m_n=0$; in particular, $m_0(y,\tau)=\int_{-\infty}^{\infty}C(\xi,y,\tau)\,\dd\xi$ is the local zeroth axial moment. The wall conditions are inherited from Eq.~\eqref{eq:23}, and the hierarchy uses the same moving coordinate as Eq.~\eqref{eq:yielding_pe}. The moment equations are solved separately from the full advection--diffusion equation; the initial variance $\sigma_0^2$ is included in comparing the initially Gaussian tracer pulse with the impulse-based formulation. Agreement of the two methods provides a direct test of finite-time axial spreading.

\begin{table*}
\caption{Time-step convergence of the direct two-dimensional solute solver against the independent finite-volume moment hierarchy at $Y=0$, $Ha_p=0.8$, $\tau=1$, $\sigma_0=0.4$, $N_y=65$ and $N_x=320$. The reference variance including the initial Gaussian pulse width is $2.96976039120$; the final column is the relative variance difference in percent. Here $N_y$ and $N_x$ denote transverse and axial grid-point counts and $\Delta\tau$ the numerical time-step size; both grid sizes are fixed.}
\label{tab:direct_dt_verification_new}
\centering
\begin{tabular}{rrrr}
\toprule
Maximum $\Delta\tau$ & Direct PDE variance & Absolute variance difference & Relative difference (\%) \\
\midrule
$0.05000$ & $3.00152978$ & $0.03176939$ & $1.06976$ \\
$0.02500$ & $2.97842620$ & $0.00866581$ & $0.29180$ \\
$0.01250$ & $2.97207096$ & $0.00231057$ & $0.07780$ \\
$0.00625$ & $2.97036556$ & $0.00060517$ & $0.02038$ \\
\bottomrule
\end{tabular}
\end{table*}

The PDE--moment variance difference in Table~\ref{tab:direct_dt_verification_new} falls by nearly a factor of four upon each halving of the time step. The observed orders, $1.87$, $1.91$ and $1.93$, approach the second-order accuracy expected for Strang splitting. At the finest step, the relative variance difference is $0.0204\%$, with mass-conservation error below $9\times10^{-14}$. These results establish time-step convergence for the tested case at the stated spatial resolution. At the same spatial resolution and a maximum time step of $0.025$, an additional independent PDE--moment comparison at $\tau=1$ gives axial variances $2.191333$ and $2.190619$ for $Y=0.4$, and $2.165210$ and $2.165002$ for $Y=0.7$, respectively. The corresponding relative differences are $0.0326\%$ and $0.0096\%$. These yielded-flow checks complement the Newtonian time-step study without implying a complete grid--time convergence analysis for every parameter combination.

\subsection{Electrolyte constraints and dimensional feasibility}\label{sec:electromagnetic_constraints}
The electromagnetic parameters in Eqs.~\eqref{eq:realparams} are convenient for isolating the response of the coupled Casson equations. In an electrolyte, however, conductivity and Debye length are not generally independent. For an ideal dilute $z:z$ electrolyte, let $n_0$ be the bulk number density of each ion species and $D_+$ and $D_-$ their ionic diffusivities. The Nernst--Einstein relation and Debye--H\"uckel screening length give~\cite{Rodrigo2024Ionic,Ng2013Slit}
\begin{equation}\label{eq:electrolyte_properties}
\sigma=\frac{n_0z^2e^2(D_++D_-)}{k_BT_{av}},\qquad
\lambda_D^2=\frac{\epsilon k_BT_{av}}{2n_0z^2e^2}.
\end{equation}
Defining the arithmetic mean ionic diffusivity $\overline D_i=(D_++D_-)/2$, elimination of $n_0$ yields
\begin{equation}\label{eq:Ha_kappa_relation}
\sigma\lambda_D^2=\epsilon\overline D_i,\qquad
\frac{Ha_p}{\kappa}=B_0\sqrt{\frac{\epsilon\overline D_i}{\mu_p}}.
\end{equation}
This relation remains valid for unequal ionic diffusivities within the ideal Nernst--Einstein approximation, but need not apply quantitatively to concentrated or strongly correlated electrolytes. It also shows that varying salt concentration at fixed $B_0$, $h$, $\epsilon$, $\mu_p$ and temperature moves $Ha_p$ and $\kappa$ together.

A dimensional comparison illustrates the restriction. We use $\epsilon=\epsilon_r\epsilon_0$ and take a representative water-like relative permittivity $\epsilon_r=78.3$ at $25\,^{\circ}\mathrm C$~\cite{Fernandez1997Permittivity}, the CODATA vacuum permittivity $\epsilon_0=8.8541878188\times10^{-12}\,\mathrm{F\,m^{-1}}$~\cite{Mohr2025CODATA}, and the dilute-ion diffusivities $D_+=1.33\times10^{-9}$ and $D_-=2.03\times10^{-9}\,\mathrm{m^2\,s^{-1}}$ for sodium and chloride ions, respectively~\cite{Rodrigo2024Ionic}. We choose $h=50\,\mu\mathrm m$ and $\mu_p=10^{-3}\,\mathrm{Pa\,s}$ \emph{as illustrative scales}, not as measured properties of a particular Casson suspension. Table~\ref{tab:physical_consistency_new} reports the ideal-electrolyte implications at $B_0=1\,\mathrm T$, and the field required to reach the smallest computed dispersion optimum, $Ha_{p,\mathrm{opt}}=0.69554$ (Table~\ref{tab:true_casson_peaks_new}).

\begin{table*}
\caption{Electrolyte-consistent dimensional examples for $h=50\,\mu\mathrm m$, $\mu_p=10^{-3}\,\mathrm{Pa\,s}$ and the water-like and ionic properties specified in the text. Each row enforces Eq.~\eqref{eq:Ha_kappa_relation}; the last column is the required magnetic field for $Ha_p=0.69554$, not a practical operating recommendation.}
\label{tab:physical_consistency_new}
\centering
\begin{tabular}{lrrrrr}
\toprule
Specified property & $\kappa$ & $\lambda_D$ (nm) & $\sigma$ (S/m) & $Ha_p$ at $1\,\mathrm T$ & $B_0$ for $Ha_p=0.69554$ (T)\\
\midrule
$\kappa=10$ & $10$ & $5000$ & $4.66\times10^{-8}$ & $3.41\times10^{-7}$ & $2.04\times10^6$\\
$\lambda_D=10\,\mathrm{nm}$ & $5000$ & $10$ & $1.16\times10^{-2}$ & $1.71\times10^{-4}$ & $4.08\times10^3$\\
$\sigma=1\,\mathrm{S/m}$ & $4.63\times10^4$ & $1.08$ & $1$ & $1.58\times10^{-3}$ & $4.40\times10^2$\\
\bottomrule
\end{tabular}
\end{table*}

The first row corresponds to the formal $\kappa=10$ parameter slice used in the numerical optima and illustrates its extreme magnetic-field requirement within the ideal aqueous model. The second row imposes a much thinner, $10$-nm EDL in the same channel. The third row reproduces a conductivity-based estimate of $Ha_p\simeq1.58\times10^{-3}$ at $1\,\mathrm T$, but its inferred value of $\kappa$ is approximately $4.63\times10^4$, not $10$. The extreme dilution in the first row and the elevated conductivity in the third may also challenge the assumptions of an ideal binary electrolyte and equilibrium Debye--H\"uckel screening. These calculations are order-of-magnitude assessments; actual Casson suspensions may have different $\epsilon$, $\mu_p$, ionic mobilities and electrical boundary conditions.

\begin{figure}
\centering
\includegraphics[width=0.73\linewidth]{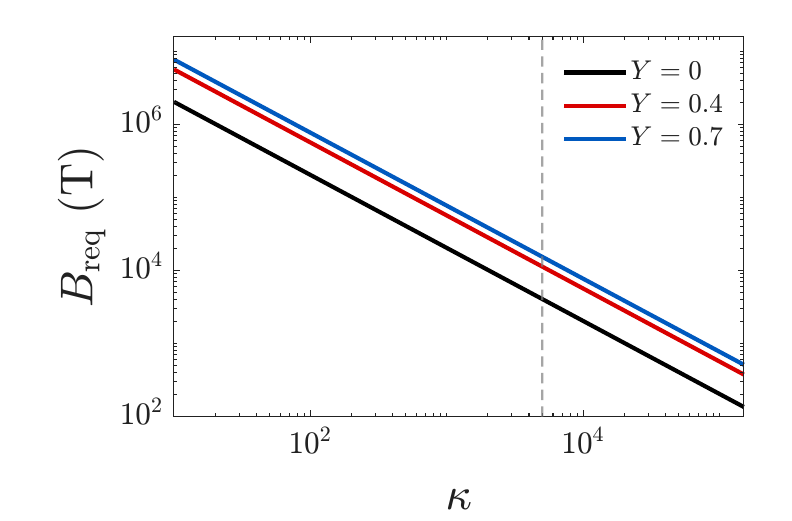}
\caption{Magnetic field $B_{\mathrm{req}}$ needed to reproduce the reference optimum Hartmann numbers obtained at $\kappa_{\mathrm{ref}}=10$, shown as a function of $\kappa$ for $Y=0$, $0.4$ and $0.7$. No optimization is repeated as $\kappa$ changes. The calculation uses Eq.~\eqref{eq:required_field} and the illustrative aqueous properties of Table~\ref{tab:physical_consistency_new}; the dashed line marks $\kappa=5000$ ($\lambda_D=10\,\mathrm{nm}$ for $h=50\,\mu\mathrm m$).}
\label{fig:field_requirement_new}
\end{figure}

Figure~\ref{fig:field_requirement_new} plots the field $B_{\mathrm{req}}$ required to reproduce the \emph{reference} optimum Hartmann numbers calculated at $\kappa_{\mathrm{ref}}=10$. Specifically,
\begin{equation}\label{eq:required_field}
B_{\mathrm{req}}(\kappa;Y)=
\frac{Ha_{p,\mathrm{opt}}(Y;\kappa_{\mathrm{ref}}=10)}{\kappa}
\sqrt{\frac{\mu_p}{\epsilon\overline D_i}}.
\end{equation}
As $\kappa$ varies in this dimensional conversion, the flow and dispersion have \emph{not} been reoptimized at each new $\kappa$; $B_{\mathrm{req}}$ is therefore a field-requirement curve for fixed target Hartmann numbers, not an independently determined $B_{\mathrm{opt}}(\kappa)$. The calculated optima lie far outside ordinary magnetic-field ranges for the representative dilute aqueous channel. At weaker fields the corresponding dispersion change may be much smaller. Applying the model to a specific device also requires its electrical current path and an assessment of Joule heating and electrochemical effects.

\subsection{Sensitivity to transverse Casson shear}\label{sec:transverse_sensitivity}
The axial model neglects the spanwise motion induced by the imposed fields. We assess the effect of this approximation by also solving a fully developed, two-component velocity problem, $\mathbf U=u^*(y^*)\mathbf e_x+w^*(y^*)\mathbf e_z$, under two specified transverse pressure closures. For shear tractions $\tau_{xy}$ and $\tau_{zy}$, the vector Casson law is defined in terms of $T=(\tau_{xy}^2+\tau_{zy}^2)^{1/2}$: when $T\leq\tau_y$ both shear rates vanish, whereas for $T>\tau_y$,
\begin{equation}\label{eq:vector_casson_test}
\begin{pmatrix}\dd u^*/\dd y^*\\[2pt]\dd w^*/\dd y^*\end{pmatrix}
=\frac{\bigl(\sqrt{T}-\sqrt{\tau_y}\bigr)^2}{\mu_p T}
\begin{pmatrix}\tau_{xy}\\[2pt]\tau_{zy}\end{pmatrix}.
\end{equation}
The calculation retains the spanwise electric-double-layer force $-\rho_e(y^*)E_z$ and the spanwise Lorentz contribution $\sigma B_0E_x-\sigma B_0^2w^*$. One closure chooses a constant transverse pressure gradient to balance the spatially uniform term $\sigma B_0E_x$; the other determines that pressure gradient from $\langle w^*\rangle=0$. Both retain no slip at the channel walls. Neither prescribes the electrode geometry or establishes the current distribution of a finite-width device.

For a sensitivity test, we use the fixed-forcing parameters $\Pi=1$, $\Omega=0.12$, $S_p=1$, $\kappa=10$ and $Pe_0=10$, with $Y=0,0.2,0.4,0.7$ and $0\leq Ha_p\leq4$. Ideal-electrolyte properties with an illustrative $|\zeta|=5\,\mathrm{mV}$ give $|E_z/E_x|=0.08464$ and $|ze\zeta/(k_BT_{av})|=0.195$. Table~\ref{tab:vector_casson_sensitivity} reports the largest deviations from the axial model for the zero-mean-spanwise-flow closure over 41 sampled Hartmann numbers, together with differences in the refined magnetic optimum. The relative enhancement error is evaluated against $D_T^{\mathrm C}-1$, not against the molecular-diffusion contribution. The alternative uniform-force-balanced closure gives smaller maximum absolute changes in $D_T^{\mathrm C}$, no larger than $2.69\times10^{-6}$ for the same cases.

\begin{table*}
\caption{Sensitivity of the normalized Casson dispersion coefficient to transverse shear under the zero-mean-spanwise-flow closure, for the $|\zeta|=5\,\mathrm{mV}$ illustrative electrolyte. Values are maxima over $0\leq Ha_p\leq4$ unless identified as optimum shifts. Here $\Delta$ denotes the two-component result minus the axial-model result, and $y_f$ denotes the upper-half-channel yield-front location. The $Y=0$ differences are at the numerical-error level.}
\label{tab:vector_casson_sensitivity}
\centering
\begin{tabular}{lrrrr}
\toprule
$Y$ & $\max|\Delta D_T^{\mathrm C}|$ & $\max|\Delta(D_T^{\mathrm C}-1)|/(D_T^{\mathrm C}-1)$ (\%) & $\max|\Delta y_f|$ & $|\Delta Ha_{p,\mathrm{opt}}|$\\
\midrule
$0$   & $<10^{-9}$              & $<10^{-5}$ & $0$                     & $<10^{-8}$\\
$0.2$ & $1.14\times10^{-5}$      & $0.0852$  & $1.42\times10^{-4}$    & $5.37\times10^{-5}$\\
$0.4$ & $6.44\times10^{-6}$      & $0.1946$  & $4.15\times10^{-4}$    & $8.31\times10^{-5}$\\
$0.7$ & $2.98\times10^{-6}$      & $0.2712$  & $7.01\times10^{-4}$    & $1.07\times10^{-4}$\\
\bottomrule
\end{tabular}
\end{table*}

The two-component solutions preserve the movement and attenuation of the magnetic dispersion maximum as $Y$ increases. At $Ha_p=1$, the fixed-throughput ordering likewise persists under both transverse-pressure closures. The magnetic-optimum trend and the fixed-throughput ordering are therefore insensitive to the small transverse shear arising under these two closures at the chosen forcing. Their applicability to a finite-width microchannel depends on the actual electrode, pressure and current conditions. The associated magnetic sweep retains the dimensional restrictions established by Eq.~\eqref{eq:Ha_kappa_relation} and Table~\ref{tab:physical_consistency_new}.

\FloatBarrier
\section{Conclusions}
We have combined a stress-resolved Casson flow calculation with Mei's multiple-scale homogenization to investigate longitudinal dispersion under pressure and electromagnetic forcing. The solvability conditions separate mean drift from dispersive spreading and yield an auxiliary problem that accounts for the velocity throughout the slit, including the unyielded core. The exact pressure-driven limit gives two distinct laws near arrest: the mean velocity decays as the third power of the remaining yielded-layer thickness, whereas the dispersion enhancement decays as its eighth power. A closed-form constant-viscosity solution provides a separate reference for the effects of the imposed fields.

The principal finding is that the effect of yielding depends on the operating constraint. With pressure and electrical forcing fixed, increasing $Y$ from $0$ to $0.7$ moves the magnetic dispersion optimum from $Ha_p=0.696$ to $2.634$, while the peak $D_T^{\mathrm C}$ decreases from $1.454$ to $1.015$. The unyielded region grows and the velocity departures responsible for dispersion diminish. At a prescribed mean velocity $\langle\widehat u\rangle=0.08$, the required pressure gradient instead adjusts to the constitutive response. At $Ha_p=1$, $D_T^{\mathrm C}$ increases from $1.000433$ to $1.002311$ over the same yield-stress range, corresponding to a factor of about $5.34$ in the shear-induced enhancement. Thus, the two operating modes exhibit opposite dispersion trends for the states examined.

This difference is explained by the energy identity obtained from the homogenized auxiliary problem. An unyielded region can move at a velocity different from the sectional mean and contribute to spreading despite having no local shear. Predicting dispersion therefore requires the complete velocity profile, not only the fraction of fluid that remains unyielded. Analytical benchmarks, mesh refinement, direct two-dimensional transport and an independent moment calculation support the velocity and dispersion predictions. The higher-order concentration reconstruction further resolves the transverse structure of the developing tracer plume.

The operating-constraint comparison provides a basis for estimating sample-band spreading in idealized microchannels carrying yield-stress suspensions under pressure or flow-rate control. The dimensional electrolyte analysis places the magnetic optima in a theoretical forcing regime for the illustrative aqueous channel. Coupled axial and spanwise Casson calculations preserve the principal trends under the two tested transverse-pressure closures in the weak-shear regime. Application to a particular device will depend on its electrode geometry, current path and material properties.

\section*{Acknowledgments}
\textbf{S.C.} acknowledges funding from the Science and Engineering Research Board (SERB), Government of India, through the Sir J. C. Bose National Fellowship.

		\section*{Declaration of interests} 
The authors declare that they have no competing interests.

		\section*{Data availability} 
No experimental data were collected. The numerical tables and plotted results were generated from the analytical formulae and MATLAB calculations described in this article. The numerical data and custom MATLAB scripts underlying the reported tables and figures are available from the corresponding author upon reasonable request.

        \appendix
\section{Multiple-scale derivation}\label{app:multiscale}
\subsection{First-order solvability condition}\label{app:firstorder}

After the exponentially decaying transverse fast modes have relaxed, the first-order correction is independent of the fast variable $t_0$, and Eq.~\eqref{eq:34} reduces to
\begin{equation}\label{eq:36}
	\frac{\partial C_{0}}{\partial t_1}+uPe\frac{\partial C_{0}}{\partial x}=\frac{\partial^2 {C_{1}}}{\partial y^2}.
\end{equation}

Averaging Eq.~\eqref{eq:36} over the channel and applying Eq.~\eqref{eq:35} gives
\begin{equation}\label{eq:38}
	\frac{\partial C_{0}}{\partial t_1}+\langle u\rangle Pe\frac{\partial C_{0}}{\partial x}=0.
\end{equation}
Subtracting the mean equation~\eqref{eq:38} and using the independence of $C_0$ from $y$ yields
\begin{equation}\label{eq:39a}
	Pe\left(u-\langle u\rangle\right)\frac{\partial C_{0}}{\partial x}=\frac{\partial^2 C_{1}}{\partial y^2}
\end{equation}

\subsection{Second-order concentration correction}\label{app:secondorder}

After relaxation of the fast transverse modes, the quasi-steady second-order correction obeys the reduction of Eq.~\eqref{eq:44}:
\begin{equation}\label{eq:46}
	\frac{\partial C_{0}}{\partial t_2}+\frac{\partial {C_{1}}}{\partial t_1}+uPe{\frac{\partial C_{1}}{\partial x}}=\frac{\partial^2 C_{0}}{\partial x^2}+\frac{\partial^2 {C_{2}}}{\partial y^2}
\end{equation}
Averaging Eq.~\eqref{eq:46} over $y$ with the wall conditions~\eqref{eq:45} gives
\begin{equation}\label{eq:47}
	\frac{\partial C_{0}}{\partial t_2}+\frac{\partial \left<{C_{1}}\right>}{\partial t_1}+Pe\left<{u\frac{\partial C_{1}}{\partial x}}\right>=\frac{\partial^2 C_{0}}{\partial x^2}
\end{equation}

Subtracting the cross-sectional mean~\eqref{eq:47} from Eq.~\eqref{eq:44} yields
\begin{equation}\label{eq:48}
	\frac{\partial C_{2}}{\partial t_0}+\frac{\partial C_{1}}{\partial t_1}+Pe{u\frac{\partial C_{1}}{\partial x}}-Pe\left<{u\frac{\partial C_{1}}{\partial x}}\right>-\frac{\partial \left<{C_{1}}\right>}{\partial t_1}=\frac{\partial^2 C_{2}}{\partial y^2}.
\end{equation}
Using Eqs.~\eqref{eq:38}, \eqref{eq:40}, \eqref{eq:43} and \eqref{eq:16}, the terms in Eq.~\eqref{eq:48} become 
\begin{align}
	&\frac{\partial \left<{C_{1}}\right>}{\partial t_1}=0\label{eq:49}&\\& \frac{\partial C_{1}}{\partial t_1}=-Pe^2\langle u\rangle A_1\frac{\partial^2 C_{0}}{\partial x^2}\label{eq:50}&\\&
	{u\frac{\partial C_{1}}{\partial x}}=Pe\,uA_1\frac{\partial^2 C_{0}}{\partial x^2}\label{eq:51}&\\&
	\left<{u\frac{\partial C_{1}}{\partial x}}\right>=Pe\langle uA_1\rangle\frac{\partial^2 C_{0}}{\partial x^2}\label{eq:52}.&
\end{align}

Substitution of Eqs.~\eqref{eq:49}--\eqref{eq:52} into Eq.~\eqref{eq:48} gives
\begin{align}\label{eq:53a}
	\frac{\partial C_{2}}{\partial t_0}+	Pe^2\left[uA_1-\langle u\rangle A_1-\langle uA_1\rangle\right]\frac{\partial^2 C_{0}}{\partial x^2}=\frac{\partial^2 C_{2}}{\partial y^2}.	
\end{align}

\subsection{Third-order concentration correction}\label{app:thirdorder}

Under the corresponding quasi-steady transverse approximation, the third-order correction of Eq.~\eqref{eq:58} satisfies
\begin{equation}\label{eq:60}
	\frac{\partial C_{1}}{\partial t_2}+\frac{\partial {C_{2}}}{\partial t_1}+uPe{\frac{\partial C_{2}}{\partial x}}=\frac{\partial^2 C_{1}}{\partial x^2}+\frac{\partial^2 {C_{3}}}{\partial y^2}.
\end{equation}
Averaging Eq.~\eqref{eq:60} across the slit gives
\begin{equation}\label{eq:61}
	\frac{\partial \left<{C_{1}}\right>}{\partial t_2}+\frac{\partial \left<{C_{2}}\right>}{\partial t_1}+Pe\left<u\frac{\partial C_2}{\partial x}\right>=\frac{\partial^2 \left<{C_{1}}\right>}{\partial x^2}.
\end{equation}
Subtracting the mean relation~\eqref{eq:61} from Eq.~\eqref{eq:58} gives
\begin{equation}\label{eq:62}
	\frac{\partial C_{3}}{\partial t_0}+\frac{\partial C_{1}}{\partial t_2}+\frac{\partial {C_{2}}}{\partial t_1}+Pe\left(u{\frac{\partial C_{2}}{\partial x}}-\left<u{\frac{\partial C_{2}}{\partial x}}\right>\right)-	\frac{\partial \left<{C_{1}}\right>}{\partial t_2}-\frac{\partial \left<{C_{2}}\right>}{\partial t_1}=\frac{\partial^2 C_{1}}{\partial x^2}+\frac{\partial^2 {C_{3}}}{\partial y^2}-\frac{\partial^2 \left<{C_{1}}\right>}{\partial x^2}.	
\end{equation}
Using Eqs.~\eqref{eq:38}, \eqref{eq:40}, \eqref{eq:43}, \eqref{eq:47}, \eqref{eq:52}, \eqref{eq:54}, \eqref{eq:57} and \eqref{eq:16}, the terms in Eq.~\eqref{eq:62} are 
\begin{align}
	&\frac{\partial \left<{C_{1}}\right>}{\partial t_2}=0\label{eq:63}&\\& \frac{\partial \left<{C_{2}}\right>}{\partial t_1}=0\label{eq:64}&\\& \frac{\partial^2 \left<{C_{1}}\right>}{\partial x^2}=0\label{eq:65}&\\& \frac{\partial C_{1}}{\partial t_2}=Pe\,A_1\left[1-Pe^2\langle uA_1\rangle\right]\frac{\partial^3 C_{0}}{\partial x^3}\label{eq:66}&\\&\frac{\partial C_{2}}{\partial t_1}=-Pe^3\langle u\rangle A_2\frac{\partial^3 C_{0}}{\partial x^3}\label{eq:67}&\\&
	{u\frac{\partial C_{2}}{\partial x}}=Pe^2\,uA_2\frac{\partial^3 C_{0}}{\partial x^3}\label{eq:68}&\\&
	\left<{u\frac{\partial C_{2}}{\partial x}}\right>=Pe^2\left<uA_2\right>\frac{\partial^3 C_{0}}{\partial x^3}\label{eq:69}.&
\end{align}

Substituting \eqref{eq:63}-\eqref{eq:69} in \eqref{eq:62}, we get
\begin{align}\label{eq:70a}
	\frac{\partial C_{3}}{\partial t_0}+	Pe^3\left[uA_2-\langle u\rangle A_2-\left<uA_2\right>-\langle uA_1\rangle A_1\right]\frac{\partial^3 C_0}{\partial x^3}=\frac{\partial^2 C_3}{\partial y^2}.
\end{align}

\subsection{Homogenized transport equation}\label{app:homogenization}

Substitution of Eqs.~\eqref{eq:49} and \eqref{eq:52} into Eq.~\eqref{eq:47} yields

\begin{align}\label{eq:75}
	\frac{\partial C_{0}}{\partial t_2}=(1-Pe^2\langle uA_1\rangle)\frac{\partial^2 C_{0}}{\partial x^2}
\end{align} 
Multiplying Eq.~\eqref{eq:38} by $\varepsilon$ and Eq.~\eqref{eq:75} by $\varepsilon^2$, and then adding, gives
\begin{align}\label{eq:76}
	\frac{\partial C_0}{\partial t_0}+\varepsilon\frac{\partial C_0}{\partial t_1}+\varepsilon^2\frac{\partial C_0}{\partial t_2}+\varepsilon Pe\langle u\rangle\frac{\partial C_0}{\partial x}=\varepsilon^2\left[1-Pe^2\langle uA_1\rangle\right]\frac{\partial^2 C_0}{\partial x^2}
\end{align}

\section{Analytical dispersion formula for the linear-viscosity model}\label{app:coefficients}
For the linear effective-viscosity velocity~\eqref{eq:16}, introduce $r_\beta=1+1/\beta$ and
\begin{equation}\label{eq:app_linear_coefficients}
\begin{aligned}
d_1&=\frac{\beta_1+S Ha}{r_\beta m^2},\qquad
 d_2=\frac{\kappa^2}{r_\beta(\kappa^2-m^2)},\\
d_3&=\langle u\rangle
=d_1\left(1-\frac{\tanh m}{m}\right)
+d_2\left(\frac{\tanh m}{m}-\frac{\tanh\kappa}{\kappa}\right),\\
d_4&=d_2\left(\frac{1}{\kappa^3}-\frac{1}{6\kappa}\right)\tanh\kappa
-(d_2-d_1)\left(\frac{1}{m^3}-\frac{1}{6m}\right)\tanh m.
\end{aligned}
\end{equation}
These coefficients yield the explicit first-order solution in the main text, including its zero-mean normalization. Its derivative may be written
\begin{equation}\label{eq:app_linear_gradient}
\frac{\dd A_1}{\dd y}
=b_0y+b_1\sinh(my)+b_2\sinh(\kappa y),
\end{equation}
where $b_0=d_1-d_3$, $b_1=(d_2-d_1)/(m\cosh m)$ and $b_2=-d_2/(\kappa\cosh\kappa)$. Hence the shear-induced factor has the exact nonnegative representation
\begin{equation}\label{eq:app_linear_factor}
D_T^*=-\langle u A_1\rangle
=\int_0^1\left[b_0 y+b_1\sinh(my)+b_2\sinh(\kappa y)\right]^2\,\dd y.
\end{equation}
For completeness, define $I_1(a)=\cosh(a)/a-\sinh(a)/a^2$,
$I_2(a)=\sinh(2a)/(4a)-1/2$ and
$I_3(a,b)=\bigl[\sinh(a+b)/(a+b)-\sinh(a-b)/(a-b)\bigr]/2$.
Direct integration gives
\begin{align}\label{eq:app_linear_factor_explicit}
D_T^*={}&\frac{b_0^2}{3}+2b_0b_1I_1(m)+2b_0b_2I_1(\kappa)
+b_1^2I_2(m)\nonumber\\
&+b_2^2I_2(\kappa)+2b_1b_2I_3(m,\kappa).
\end{align}
Expressions at $m=0$ or $m=\kappa$ are taken in their combined continuous limits. Numerically, the integral form~\eqref{eq:app_linear_factor} or the auxiliary differential equation is preferable when cancellation makes the separated coefficients ill-conditioned.

\section{Integral representation of auxiliary boundary-value problems}\label{app:neumann_operator}
For an integrable function $f$ on $[-1,1]$ satisfying $\langle f\rangle=0$, define
\begin{align}\label{eq:green_neumann}
\mathcal G[f](y)={}&\int_{-1}^{y}(y-s)f(s)\,\dd s\nonumber\\
&-\frac12\int_{-1}^{1}\!\int_{-1}^{\eta}(\eta-s)f(s)\,\dd s\,\dd\eta.
\end{align}
Then $\mathcal G[f]$ is the unique zero-mean solution of
$\dd^2\mathcal G[f]/\dd y^2=f$ with
$\dd\mathcal G[f]/\dd y=0$ at both walls. Writing $v=u-\langle u\rangle$,
\begin{align}\label{eq:green_higher_auxiliary}
A_1&=\mathcal G[v],\nonumber\\
A_2&=\mathcal G\left[vA_1-\langle uA_1\rangle\right],\nonumber\\
A_3&=\mathcal G\left[vA_2-\langle uA_2\rangle-\langle uA_1\rangle A_1\right].
\end{align}
All three forcing functions have zero cross-sectional average because $\langle A_1\rangle=\langle A_2\rangle=0$. Thus the representation satisfies the solvability conditions, wall conditions and gauge normalizations in Eqs.~\eqref{eq:41}--\eqref{eq:43}, \eqref{eq:55}--\eqref{eq:57} and \eqref{eq:72}--\eqref{eq:74}, without introducing unwieldy coefficient expansions. The same operator provides the Casson auxiliary solutions when $u$ is replaced by $\widehat u$ and the corresponding velocity-scale convention is used.

\section{Dispersion near cessation of pressure-driven flow}\label{app:near_arrest}
The calculation follows from the exact velocity profile \eqref{eq:casson_exact}, the mean-velocity formula \eqref{eq:casson_mean}, and the Taylor--Aris energy identity \eqref{eq:positive_dispersion}. For the pressure-driven Casson limit, consider the upper half-channel $y\in[0,1]$ and $\alpha=1-\delta$ with $0<\delta\ll1$. The mean velocity follows from Eq.~\eqref{eq:casson_mean} as $\langle u^*\rangle/U_P=\int_\alpha^1 r(\sqrt r-\sqrt\alpha)^2\,\dd r$. Substitution of $r=\alpha+\delta s$ gives $\langle u^*\rangle/U_P=\delta^3/12+O(\delta^4)$. Inside the plug, $0\leq y<\alpha$, the velocity is constant and equals $U_PF(\alpha)$. Expanding Eq.~\eqref{eq:casson_F} then gives
\begin{equation}\label{eq:app_plug_difference}
\frac{u_{\rm plug}}{U_P}-\frac{\langle u^*\rangle}{U_P}
=\frac{\delta^4}{48}+O(\delta^5).
\end{equation}
Let $a(y)$ solve $\dd^2a/\dd y^2=u^*(y)/U_P-\langle u^*\rangle/U_P$ with $\dd a/\dd y=0$ at $y=0$ and $y=1$. Throughout the plug, $\dd a/\dd y=\delta^4 y/48+O(\delta^5)$; the yielded layers have width $\delta$ and contribute only at higher order to the auxiliary-function energy. Using reflection symmetry,
\begin{align}\label{eq:app_dispersion_scaling}
\frac{D_{\rm eff}}{D}-1
&=\left(\frac{U_Ph}{D}\right)^2
\int_0^1\left(\frac{\dd a}{\dd y}\right)^2\,\dd y\\
&=\left(\frac{U_Ph}{D}\right)^2
\left[\frac{\delta^8}{48^2}\int_0^1y^2\,\dd y
+O(\delta^9)\right]\\
&=\left(\frac{U_Ph}{D}\right)^2
\left[\frac{\delta^8}{6912}+O(\delta^9)\right].
\end{align}
The prefactor pertains to the half-height velocity scale $U_P=Gh^2/\mu_p$, and should not be transferred unchanged to alternative velocity scales, geometries, fixed-flux constraints, or spatially variable tracer diffusivities.

\bibliography{reference}
\end{document}